\documentclass[a4paper,11pt]{article}
\pdfoutput=1
\usepackage{jheppubMod}
\usepackage{amsmath,amsfonts,amssymb,array}
\usepackage{bm,bbm,bbm,bbold}
\usepackage{datetime,dsfont}
\usepackage{enumerate}
\usepackage{float,fancyvrb}
\usepackage[T1]{fontenc}
\usepackage{graphicx}
\usepackage{lineno,lipsum,longtable}
\usepackage{multirow,multicol}
\usepackage{subfigure,slashed}
\usepackage{tabu}
\usepackage{pdflscape}
\usepackage{soul}
\usepackage[toc,page]{appendix}
\usepackage{url}
\usepackage{diagbox}
\usepackage[utf8]{inputenc}
\usepackage[T1]{fontenc}
\usepackage{cancel}
\usepackage{tikz-feynman,contour}
\tikzfeynmanset{compat=1.1.0}
\usepackage{mathtools}
\usepackage{environ}
\usepackage{caption}
\usepackage{comment}
\usepackage{xspace}

\usepackage{epstopdf}
\hypersetup{bookmarks=true,unicode=true,pdftoolbar=true,pdfmenubar=true,
  pdffitwindow=false,pdfstartview={FitH},
  pdftitle={Title},pdfauthor={Names},
  pdfsubject={Subject},pdfcreator={Creator},pdfproducer={Producer},
  pdfkeywords={Helicity Polarization}{Monte Carlo Simulations}{Vector Boson Scattering}{Composite Higgs},
  pdfnewwindow=true,colorlinks=true
}

\newcolumntype{x}[1]{
{\centering\hspace{0pt}}p{#1}}

\newcommand{\bmp}{\noindent\begin{minipage}{16cm}}
\newcommand{\emp}{\end{minipage}\vskip 7mm} 

\newcommand{\GeV}{\mbox{ ${\mathrm{GeV}}$}}
\newcommand{\TeV}{\mbox{ ${\mathrm{TeV}}$}}
\newcommand{\ifb}{\mbox{ ${\mathrm{fb^{-1}}}$}}

\newcommand{\fb}{\mbox{ ${\mathrm{fb}}$}}

\newcommand*{\pT}{\ensuremath{p_{\text{T}}}\xspace}

\newcommand*{\HT}{\ensuremath{H_{\text{T}}}\xspace}
\newcommand*{\met}{\ensuremath{E_{\text{T}}^{\text{miss}}}\xspace}

\newcommand{\mg}{\textsc{MG5\_aMC@NLO}\xspace}

\newcommand{\pythia}{\textsc{Pythia8}\xspace}
\newcommand{\delphes}{\textsc{Delphes}\xspace}
\newcommand{\fastjet}{\textsc{Fastjet}\xspace}

\newcommand{\SRggg}{\ensuremath{\texttt{SR}^{\texttt{3}\gamma}}\xspace}
\newcommand{\SRgg}{\ensuremath{\texttt{SR}^{\gamma\gamma}}\xspace}
\newcommand{\SRWH}{\ensuremath{\texttt{SR}^{\gamma\gamma}_{\texttt{W-H}}}\xspace}

\def\fig#1{{Fig.~\ref{#1}}}
\def\figs#1#2{{Figs.~\ref{#1}--\ref{#2}}}
\def\sec#1{{Sec.~\ref{#1}}}
\def\tab#1{{Tab.~\ref{#1}}}

\newcommand{\beq}{\begin{equation}}
\newcommand{\eeq}{\end{equation}}
\newcommand{\beqs}{\begin{eqnarray}}
\newcommand{\eeqs}{\end{eqnarray}}
\newcommand{\bit}{\begin{itemize}}
\newcommand{\eit}{\end{itemize}}
\newcommand{\bce}{\begin{center}}
\newcommand{\ece}{\end{center}}
\newcommand{\ben}{\begin{enumerate}}
\newcommand{\een}{\end{enumerate}}

\newcommand{\hc}{\mathrm{h.c.}}

\newcommand{\nn}{\nonumber}

\newcommand*{\chinoonep}{\ensuremath{\mathchoice%
      {\displaystyle\raise.4ex\hbox{\(\displaystyle\tilde\chi^+_1\)}}%
         {\textstyle\raise.4ex\hbox{\(\textstyle\tilde\chi^+_1\)}}%
       {\scriptstyle\raise.3ex\hbox{\(\scriptstyle\tilde\chi^+_1\)}}%
 {\scriptscriptstyle\raise.3ex\hbox{\(\scriptscriptstyle\tilde\chi^+_1\)}}}\xspace}
 
\newcommand*{\chinoonem}{\ensuremath{\mathchoice%
      {\displaystyle\raise.4ex\hbox{\(\displaystyle\tilde\chi^-_1\)}}%
         {\textstyle\raise.4ex\hbox{\(\textstyle\tilde\chi^-_1\)}}%
       {\scriptstyle\raise.3ex\hbox{\(\scriptstyle\tilde\chi^-_1\)}}%
 {\scriptscriptstyle\raise.3ex\hbox{\(\scriptscriptstyle\tilde\chi^-_1\)}}}\xspace}

\newcommand*{\chinoonepm}{\ensuremath{\mathchoice%
      {\displaystyle\raise.4ex\hbox{\(\displaystyle\tilde\chi^\pm_1\)}}%
         {\textstyle\raise.4ex\hbox{\(\textstyle\tilde\chi^\pm_1\)}}%
       {\scriptstyle\raise.3ex\hbox{\(\scriptstyle\tilde\chi^\pm_1\)}}%
 {\scriptscriptstyle\raise.3ex\hbox{\(\scriptscriptstyle\tilde\chi^\pm_1\)}}}\xspace}

\newcommand*{\chinotwop}{\ensuremath{\mathchoice%
      {\displaystyle\raise.4ex\hbox{\(\displaystyle\tilde\chi^+_2\)}}%
         {\textstyle\raise.4ex\hbox{\(\textstyle\tilde\chi^+_2\)}}%
       {\scriptstyle\raise.3ex\hbox{\(\scriptstyle\tilde\chi^+_2\)}}%
 {\scriptscriptstyle\raise.3ex\hbox{\(\scriptscriptstyle\tilde\chi^+_2\)}}}\xspace}
 
\newcommand*{\chinotwom}{\ensuremath{\mathchoice%
      {\displaystyle\raise.4ex\hbox{\(\displaystyle\tilde\chi^-_2\)}}%
         {\textstyle\raise.4ex\hbox{\(\textstyle\tilde\chi^-_2\)}}%
       {\scriptstyle\raise.3ex\hbox{\(\scriptstyle\tilde\chi^-_2\)}}%
 {\scriptscriptstyle\raise.3ex\hbox{\(\scriptscriptstyle\tilde\chi^-_2\)}}}\xspace}

\newcommand*{\chinotwopm}{\ensuremath{\mathchoice%
      {\displaystyle\raise.4ex\hbox{\(\displaystyle\tilde\chi^\pm_2\)}}%
         {\textstyle\raise.4ex\hbox{\(\textstyle\tilde\chi^\pm_2\)}}%
       {\scriptstyle\raise.3ex\hbox{\(\scriptstyle\tilde\chi^\pm_2\)}}%
 {\scriptscriptstyle\raise.3ex\hbox{\(\scriptscriptstyle\tilde\chi^\pm_2\)}}}\xspace}

\newcommand*{\ninoone}{\ensuremath{\mathchoice%
      {\displaystyle\raise.4ex\hbox{\(\displaystyle\tilde\chi^0_1\)}}%
         {\textstyle\raise.4ex\hbox{\(\textstyle\tilde\chi^0_1\)}}%
       {\scriptstyle\raise.3ex\hbox{\(\scriptstyle\tilde\chi^0_1\)}}%
 {\scriptscriptstyle\raise.3ex\hbox{\(\scriptscriptstyle\tilde\chi^0_1\)}}}\xspace}
 
\newcommand*{\ninotwo}{\ensuremath{\mathchoice%
      {\displaystyle\raise.4ex\hbox{\(\displaystyle\tilde\chi^0_2\)}}%
         {\textstyle\raise.4ex\hbox{\(\textstyle\tilde\chi^0_2\)}}%
       {\scriptstyle\raise.3ex\hbox{\(\scriptstyle\tilde\chi^0_2\)}}%
 {\scriptscriptstyle\raise.3ex\hbox{\(\scriptscriptstyle\tilde\chi^0_2\)}}}\xspace}
 
\newcommand*{\ninothree}{\ensuremath{\mathchoice%
      {\displaystyle\raise.4ex\hbox{\(\displaystyle\tilde\chi^0_3\)}}%
         {\textstyle\raise.4ex\hbox{\(\textstyle\tilde\chi^0_3\)}}%
       {\scriptstyle\raise.3ex\hbox{\(\scriptstyle\tilde\chi^0_3\)}}%
 {\scriptscriptstyle\raise.3ex\hbox{\(\scriptscriptstyle\tilde\chi^0_3\)}}}\xspace}
 
\newcommand*{\ninofour}{\ensuremath{\mathchoice%
      {\displaystyle\raise.4ex\hbox{\(\displaystyle\tilde\chi^0_4\)}}%
         {\textstyle\raise.4ex\hbox{\(\textstyle\tilde\chi^0_4\)}}%
       {\scriptstyle\raise.3ex\hbox{\(\scriptstyle\tilde\chi^0_4\)}}%
 {\scriptscriptstyle\raise.3ex\hbox{\(\scriptscriptstyle\tilde\chi^0_4\)}}}\xspace} 

\newcommand*{\ninofive}{\ensuremath{\mathchoice%
      {\displaystyle\raise.4ex\hbox{\(\displaystyle\tilde\chi^0_5\)}}%
         {\textstyle\raise.4ex\hbox{\(\textstyle\tilde\chi^0_5\)}}%
       {\scriptstyle\raise.3ex\hbox{\(\scriptstyle\tilde\chi^0_5\)}}%
 {\scriptscriptstyle\raise.3ex\hbox{\(\scriptscriptstyle\tilde\chi^0_5\)}}}\xspace}

\newcommand*{\pgldone}{\ensuremath{\mathchoice%
      {\displaystyle\hbox{\(\displaystyle\tilde G\)}}%
         {\textstyle\hbox{\(\textstyle\tilde G\)}}%
       {\scriptstyle\hbox{\(\scriptstyle\tilde G\)}}%
 {\scriptscriptstyle\hbox{\(\scriptscriptstyle\tilde G\)}}}\xspace} 

\newcommand*{\pgldtwo}{\ensuremath{\mathchoice%
      {\displaystyle\hbox{\(\displaystyle\tilde G'\)}}%
         {\textstyle\hbox{\(\textstyle\tilde G'\)}}%
       {\scriptstyle\hbox{\(\scriptstyle\tilde G'\)}}%
 {\scriptscriptstyle\hbox{\(\scriptscriptstyle\tilde G'\)}}}\xspace} 
 
 \newcommand*{\pgldthree}{\ensuremath{\mathchoice%
      {\displaystyle\hbox{\(\displaystyle\tilde G''\)}}%
         {\textstyle\hbox{\(\textstyle\tilde G''\)}}%
       {\scriptstyle\hbox{\(\scriptstyle\tilde G''\)}}%
 {\scriptscriptstyle\hbox{\(\scriptscriptstyle\tilde G''\)}}}\xspace} 

 \newcommand*{\pgldi}{\ensuremath{\mathchoice%
      {\displaystyle\hbox{\(\displaystyle\tilde G^i\)}}%
         {\textstyle\hbox{\(\textstyle\tilde G^i\)}}%
       {\scriptstyle\hbox{\(\scriptstyle\tilde G^i\)}}%
 {\scriptscriptstyle\hbox{\(\scriptscriptstyle\tilde G^i\)}}}\xspace}

\begin{document}
\title{Electroweak signatures of gauge-mediated supersymmetry breaking in multiple hidden sectors}

\author[a,b]{Diogo Buarque Franzosi}
\author[a]{Gabriele Ferretti}
\author[b]{Ellen Riefel} 
\author[b]{Sara Strandberg}

\affiliation[a]{Chalmers University of Technology, Department of Physics, 
41296 G\"oteborg, Sweden }
\affiliation[b]{Stockholm University, Department of Physics,
106 91 Stockholm, Sweden}
\emailAdd{buarque@chalmers.se}
\emailAdd{ferretti@chalmers.se}
\emailAdd{ellen.riefel@fysik.su.se}
\emailAdd{strandberg@fysik.su.se}

\abstract{
This paper discusses electroweak collider signatures of the NMSSM with multiple-sector gauge mediation. We focus on the production of neutralinos and charginos which cascade decay into standard model particles and lighter supersymmetric particles, with special emphasis on final states with multiple photons.
A search strategy for signatures with at least three photons is presented and compared with current exclusion limits based on two-photon searches. We show that in many regions of the parameter space our strategy gives stronger constraints than the existing two-photon analysis for these models. 
 }

\maketitle

\section{Introduction}

Supersymmetry (SUSY) is one of the main targets in the search for physics beyond the standard model (BSM) at the Large Hadron Collider (LHC) but, so far, no evidence in its favor has been found. There are many aspects of the current experimental situation that are problematic for SUSY. The first, most obvious, is the absence of direct detection of superpartners, pushing some of their masses well beyond the electroweak (EW) scale. Secondly, SUSY breaking must not introduce excessive neutral flavor-changing interactions. Thirdly, the discovery of the Higgs boson \cite{Aad:2012tfa,Chatrchyan:2012ufa} at 125~GeV pushes the minimal supersymmetric standard model (MSSM, see \cite{Martin:1997ns} and references therein) into a highly fine-tuned regime requiring stop masses even larger than current direct searches or large $A_t$-terms. Lastly, the lack of signal in direct dark matter searches has made the neutralino  less appealing as a dark matter candidate.
It should be kept in mind however that the severity of some of these obstacles is partly model dependent and can be mitigated by considering non-minimal models of SUSY.  In our opinion this warrants for enlarging the class of models being targeted beyond the original MSSM with the usual SUSY breaking mechanisms.

As a specific example of the approach above, we consider the $R$-parity conserving next-to-minimal supersymmetric standard model (NMSSM, see \cite{Maniatis:2009re,Ellwanger:2009dp} and references therein) coupled via gauge mediation (GM, see \cite{Giudice:1998bp} and references therein) to multiple ($n$) SUSY breaking sectors. 
The advantages of GM are that it does not suffer from excessive neutral flavor-changing processes and that it provides a dark matter candidate, the gravitino, which is not in tension with the bounds from direct dark matter searches. The reason for concentrating on the NMSSM is on the other hand to alleviate the need for large loop contributions to the Higgs mass by allowing its tree-level value to exceed the mass of the $Z$ boson. This is particularly useful when using GM as the SUSY-breaking mechanism since the $A_t$-terms generated are typically very small, which would put further stress on generating the right Higgs mass within the MSSM alone.
Our interest in studying multiple sectors of gauge-mediated supersymmetry breaking stems from their hitherto unexplored experimental signatures with multiple gauge bosons. If one leaves aside fine-tuning issues all our results apply just as well to the MSSM with multiple GM breaking sectors. 

Multiple SUSY breaking sectors were first introduced in the context of gravity mediation~\cite{Cheung:2010mc,Cheung:2010qf}; see
also~\cite{Benakli:2007zza,Craig:2010yf,McCullough:2010wf,Cheng:2010mw,Izawa:2011hi,Thaler:2011me,Cheung:2011jq,Bertolini:2011tw}.
They were then considered for GM first in~\cite{Argurio:2011hs}. As in ordinary GM, the phenomenology is mostly driven by the lightest observable sector particle (LOSP). In~\cite{Argurio:2011gu} the LOSP was taken to be a gaugino or a stau, 
in~\cite{Liu:2013sx} the case of a higgsino LOSP was considered, 
in~\cite{Ferretti:2013wya} the LOSP was a gaugino and the main production mode was via a slepton pair,
in~\cite{Hikasa:2014yra} both higgsino and gaugino LOSPs decaying to a heavy SM boson were considered and
in~\cite{Liu:2014lda} the LOSP was a gaugino and the main production mode was vector boson fusion.
For related work see also~\cite{Baryakhtar:2012rz,DHondt:2013cwd,Iyer:2014iha,Dai:2021eah}.

In this work, we focus on electroweak production of charginos and neutralinos which cascade decay into SM particles and lighter SUSY particles.
While the chargino sector (comprised of $\chinoonepm, \chinotwopm$) is identical to that of the MSSM, the neutralino sector is extended by $n$ states, one from each SUSY breaking sector. The five heaviest mass eigenstates ($\ninoone,\cdots,\ninofive$) largely coincide with the five neutralinos from the NMSSM while the $n$ lightest mass eigenstates (\pgldi) have a large overlap with the pseudo-goldstinos (PGLDs, denoted by $\tilde\eta^i$) arising from the $n$ SUSY breaking sectors. The lightest \pgldi coincides with the nearly massless helicity $\pm 1/2$ components of the gravitino. We consider the case where all sleptons and squarks are decoupled from the spectrum, resulting in the LOSP being the $\ninoone$, which is almost purely bino. The LOSP will cascade decay to the collider-stable next-to-lightest \pgldi by the emission of one or several SM bosons. In this paper we discuss different collider signatures and perform a detailed analysis for the case of multi-photon ($n_\gamma\geq 3$) final states.

The paper is organized as follows. In Sec.~\ref{sec:model} we set the theoretical basis for our analysis. We describe the EW sector of the NMSSM lagrangian coupled to multiple GM sectors and present the main collider signatures of interest.
In Sec.~\ref{sec:benchmarks} we focus on multi-photon signatures, and construct the benchmark points to be analyzed.
Sec.~\ref{sec:simulation} concerns the details of the simulation, Sec.~\ref{sec:selection} the object definition and event selection and Sec.~\ref{sec:background_estimates} the background estimates.
We present the results on the expected reach of the multiphoton analysis in Sec.~\ref{sec:results} and offer our conclusions in Sec.~\ref{conclusion}. Details on the recast of the existing ATLAS search for the two photon signal and on the validation of our analysis are collected in the appendix.

\section{Gauge-mediated NMSSM with multiple sectors}
\label{sec:model}
For definitiveness we work in the CP preserving version of the NMSSM without holomorphic linear and quadratic soft terms, characterized by a scalar potential $V=V_F + V_D + V_S$ with
\begin{align}
  V_F &= \lambda^2 |S|^2 (H_u^\dagger H_u + H_d^\dagger H_d) + | \lambda H_u \epsilon H_d + k S^2|^2 \\
  V_D &= \frac{g_2^2}{2}|H_u^\dagger H_d|^2 + \frac{g_1^2 +g_2^2}{8}(H_u^\dagger H_u -  H_d^\dagger H_d)^2 \\
  V_S &= m_{H_u}^2 H_u^\dagger H_u + m_{H_d}^2 H_d^\dagger H_d + m_S^2 |S|^2 +\left(  \lambda A_\lambda H_u \epsilon H_d S + \frac{1}{3} k A_k S^3 + \hc \right)
\end{align}
We use the conventions of~\cite{Maniatis:2009re} and refer to that paper for definitions of the quantities involved.

We choose the following parametrization of the Higgs vacuum expectation values (vevs) ($v=246$~GeV)
\beq
 \langle S \rangle = \frac{v_s}{\sqrt 2},\quad
 \langle H_u \rangle = \begin{pmatrix} 0 \\ \frac{v \sin\beta }{\sqrt 2} \end{pmatrix},\quad
 \langle H_d \rangle = \begin{pmatrix}  \frac{v \cos\beta}{\sqrt 2} \\ 0\end{pmatrix}
\eeq
and use the tadpole equations to trade the soft mass terms $m_S^2,\, m_{H_u}^2,\, m_{H_d}^2$ for the vevs $v_s,\, v$ and $\tan\beta$. The resulting expression for the mass matrices of the neutral scalar, neutral pseudo-scalar and charged Higgs bosons are given in~\cite{Maniatis:2009re}.
The bosonic sector is thus the same as the usual NMSSM case. This is also true for the chargino sector, since the only non-standard feature is the presence of additional neutral fermions from the multiple SUSY breaking sectors. Thus the chargino mass matrix is also as in the NMSSM
\beq
          {\mathcal M}^\pm = \begin{pmatrix} M_W  & \sqrt{2} m_W \sin\beta \\ \sqrt{2} m_W \cos\beta & \lambda v_s/\sqrt{2} \end{pmatrix}
\eeq
where $M_W$ is the wino soft mass and $m_W$ is the mass of the $W$ boson.
The qualitative difference comes from the structure of the generalized neutralino mass matrix, which includes the NMSSM neutral gauginos, higgsinos and singlino and one PGLD for each SUSY breaking sector. 
In the gauge basis $\tilde B, \tilde W^3, \tilde H_u, \tilde H_d, \tilde S, \tilde \eta_1 \dots \tilde \eta_n$, the full mass matrix has the form
\beq
          {\mathcal M}^0 = \begin{pmatrix}  {\mathcal M}_{\mathrm{NMSSM}} & {\mathcal M}_{\mathrm{MIX}}\\
                                                                        {\mathcal M}^\dagger_{\mathrm{MIX}}  &  {\mathcal M}_{\mathrm{PGLD}}                  
                                       \end{pmatrix}
\eeq
where 
\beq
          {\mathcal M}_{\mathrm{NMSSM}}  = 
                                       \begin{pmatrix}
                                       M_B & 0 & -m_Z \sin\theta\cos\beta & m_Z \sin\theta\sin\beta & 0 \\
                                       0 & M_W & m_Z \cos\theta\cos\beta & -m_Z \cos\theta\sin\beta & 0 \\
                                        -m_Z \sin\theta\cos\beta &  m_Z \cos\theta\cos\beta & 0 & - \lambda v_s/\sqrt 2 & - \lambda v \sin\beta /\sqrt 2\\
                                        m_Z \sin\theta\sin\beta & -m_Z \cos\theta\sin\beta &- \lambda v_s/\sqrt 2 & 0 &  - \lambda v \cos\beta /\sqrt 2\\
                                        0 & 0 &  - \lambda v \sin\beta /\sqrt 2 &  - \lambda v \cos\beta /\sqrt 2 & \sqrt 2 k v_s
                                       \end{pmatrix}
\eeq
is the usual mass matrix for the NMSSM.
To construct the mixing matrix ${\mathcal M}_{\mathrm{MIX}}$ as well as the couplings of the neutralinos with the EW gauge bosons and the Higgs sector, we use the constrained superfields~\cite{Komargodski:2009rz} (see also~\cite{Rocek:1978nb,Lindstrom:1979kq,Samuel:1982uh})
\beq
          {\mathbf X}_i = f_i \theta^2 + \sqrt 2 \tilde \eta_i \theta + \frac{\tilde \eta_i^2}{2 f_i},
\eeq
one for each SUSY breaking sector $i=1\dots n$, obeying $  {\mathbf X}_i^2=0$.
We then write each soft term as a spurionic SUSY term
\begin{align}
  {\mathcal L} \supset  &-\int {\mathrm d}^2\theta \sum_i {\mathbf X}_i \left\{ \frac{M_{B(i)}}{2 f_i} {\mathbf B}{\mathbf B} + \frac{M_{W(i)}}{2 f_i} {\mathbf W}^a{\mathbf W}^a +
    \frac{\lambda A_{\lambda(i)}}{f_i}  {\mathbf H}_u \epsilon {\mathbf H}_d {\mathbf S} + \frac{k A_{k (i)}}{3 f_i}{\mathbf S}^3 \right\}  \nn\\
    & -\int {\mathrm d}^2\theta {\mathrm d}^2\bar\theta\sum_i {\mathbf X}^\dagger_i  {\mathbf X}_i \left\{ \frac{m^2_{H_u(i)}}{f^2_i} {\mathbf H}_u^\dagger {\mathbf H}_u +
    \frac{m^2_{H_d(i)}}{f^2_i} {\mathbf H}_d^\dagger {\mathbf H}_d + \frac{m^2_{S(i)}}{f^2_i} {\mathbf S}^\dagger {\mathbf S} \right\},
    \label{spurionsusy}
\end{align}
where the normalizations have been chosen in such a way that $M_{B} = \sum_i M_{B(i)} $ and similarly for the other soft terms.
We expand to find the $5\times n$ block ${\mathcal M}_{\mathrm{MIX}}$ of the mass matrix mixing the PGLDs with the NMSSM gauginos, higgsinos and singlino:
\beq
         {\mathcal M}_{\mathrm{MIX}} = \begin{pmatrix} 
                                       -\frac{M_{B(1)}}{\sqrt 2 f_1} D_Y & \dots & -\frac{M_{B(n)}}{\sqrt 2 f_n} D_Y\\
                                      -\frac{M_{W(1)}}{\sqrt 2 f_1} D_{T^3} & \dots & -\frac{M_{W(n)}}{\sqrt 2 f_n} D_{T^3} \\
                                       -\frac{v\left(\sqrt 2 m^2_{H_d(1)} \cos\beta - \lambda A_{\lambda(1)}v_s  \sin\beta\right)}{2 f_1} &
                                     \dots &-\frac{v\left(\sqrt 2 m^2_{H_d(n)} \cos\beta - \lambda A_{\lambda(n)}v_s \sin\beta\right)}{2 f_n}\\
                                       -\frac{v\left(\sqrt 2 m^2_{H_u(1)} \sin\beta - \lambda A_{\lambda(1)}v_s  \cos\beta\right)}{2 f_1} &
                                     \dots &-\frac{v\left(\sqrt 2 m^2_{H_u(n)} \sin\beta - \lambda A_{\lambda(n)}v_s \cos\beta\right)}{2 f_n}\\
                                     -\frac{\sqrt 2 m^2_{S(1)} v_s -  \lambda A_{\lambda(1)} v^2 \sin\beta\cos\beta + k A_{k(1)} v_s^2}{2 f_1} & 
                                     \dots &  -\frac{\sqrt 2 m^2_{S(n)} v_s -  \lambda A_{\lambda(n)} v^2 \sin\beta\cos\beta + k A_{k(n)} v_s^2}{2 f_n} 
                                                              \end{pmatrix},
\eeq
with $D_Y = -\frac{1}{4} g_1 v^2 \cos2\beta$ and $D_{T^3}= +\frac{1}{4} g_2 v^2 \cos2\beta$ being the D-terms associated to the gauge groups $U(1)_Y$ and $SU(2)_L$.
Lastly, the pure PGLD contribution~\cite{Argurio:2011hs} is given by an $n\times n$ symmetric matrix obeying the constraint of having one zero eigenvalue whose eigenvector is proportional to $(f_1, f_2 \dots f_n)^T$, corresponding to the true goldstino that becomes the $\pm 1/2$ helicity component on the nearly massless gravitino. Such a matrix has the form
\beq
\mathcal{M}_{\mathrm{PGLD}} {=}
{\small
\left(
\begin{array}{cccc}
 -\frac{f_2 \mu_{12}+f_3 \mu_{13}+\cdots+f_n \mu_{1n}}{f_1} & \mu_{12} & \cdots & \mu_{1n} \\
 \mu_{12} & -\frac{f_1 \mu_{12}+f_3 \mu_{23}+\cdots+f_n \mu_{2n}}{f_2} & \cdots & \mu_{2n} \\
\vdots & \vdots & \ddots & \vdots \\
\mu_{1n} & \mu_{2n} & \cdots & -\frac{f_1 \mu_{1n}+f_2 \mu_{2n}+\cdots+f_{n-1} \mu_{n-1\,n}}{f_n}
\end{array}
\right)}~.
\label{Mnn}
\eeq
where $\mu_{ij}$ are mass parameters characterizing the mixing between the hidden sectors induced by their coupling to the NMSSM~\cite{Argurio:2011hs}.

The Lagrangian involving only NMSSM fields is left unchanged and the couplings of two or more PGLDs are subleading and therefore neglected.
The couplings in (\ref{spurionsusy}) between the NMSSM fields and the PGLDs are relevant because they mediate the exotic decays of the neutralinos after rotating to the mass eigenbasis.
In the gauge basis they read
\begin{align}
   {\mathcal{L}}\supset & \sum_i  \frac{i M_{B(i)} }{2 \sqrt{2} f_i} \tilde B \sigma^\mu\bar\sigma^\nu \tilde \eta_i B_{\mu\nu} +
    \frac{i M_{W(i)} }{2 \sqrt{2} f_i} \tilde W^a \sigma^\mu\bar\sigma^\nu \tilde \eta_i W^a_{\mu\nu} \nn  \\
    &+\left( \frac{m^2_{H_u(i)}}{f_i} \tilde \eta_i  H_u^\dagger \tilde H_u + \frac{m^2_{H_d(i)}}{f_i} \tilde \eta_i  H_d^\dagger \tilde H_d + 
    \frac{m^2_{S(i)}}{f_i} \tilde \eta_i  S^\dagger \tilde S +  \hc \right). \label{exotic}
\end{align}
 
In the phenomenological analysis we will only be interested in the $n=3$ case since 
the next-to-lightest \pgldi is collider stable. This means that, in practice, the collider signature of the $n=2$ case is very similar to the usual $n=1$ scenario explored by LHC searches~\cite{Aaboud:2018doq,CMS:2019oou} ({\it e.g.} two photons plus missing transverse momentum) but with a massive \pgldi as an ``effective'' gravitino.
We will not consider the $n\geq 4$ cases since these tend to have collider-stable \pgldi and soft photons leading to signatures that are indistinguishable from the $n=3$ case.

For ease of comparison with the standard SUSY literature we still denote the mass eigenstates aligned with the  NMSSM  neutral fermions by 
$\ninoone, \cdots, \ninofive$ and the lower mass neutralinos aligned with the PGLDs by
 $\pgldi = \pgldone, \pgldtwo, \pgldthree$. More concretely, the gauge basis  
\noindent $\tilde \Psi = (\tilde B, \tilde W^3, \tilde H_d, \tilde H_u, \tilde S, \tilde\eta_1, \tilde\eta_2, \tilde\eta_3)$ and the mass basis
 $\tilde X = (\pgldone, \pgldtwo, \pgldthree, \ninoone, \ninotwo, \ninothree, \ninofour, \ninofive)$ are related by
 $\tilde X = N\tilde\Psi$, with $\tilde\Psi^T \mathcal{M}^0 \tilde\Psi = \tilde X^T N \mathcal{M}^0 N^T \tilde X =  \tilde X^T \mathcal{M}^0_{\mathrm{ diag.}} \tilde X$.

The scenarios phenomenologically interesting that we typically obtain in our exploration of parameter space (see \sec{sec:benchmarks} for details) have, in order of increasing mass: 
a massless goldstino $\pgldone \approx \tilde\eta_1$, followed by two neutralinos that are mostly PGLDs, $\pgldtwo \approx \tilde\eta_2$ and $\pgldthree \approx \tilde\eta_3$, a mostly bino
neutralino $\ninoone \approx \tilde B$ and finally a nearly degenerate chargino and neutralino pair mostly aligned with the corresponding winos, $\chinoonepm \approx \tilde W^\pm$ and 
$\ninotwo \approx \tilde W^3$.
We do not consider the heavier charginos and neutralinos that are mostly aligned with higgsinos and singlinos. We also decouple all sfermions and the gluinos and consider a scalar sector with all scalars heavier than the $h(125)$.

\begin{figure}[htbp]
    \centering
    \includegraphics[width=0.45\textwidth]{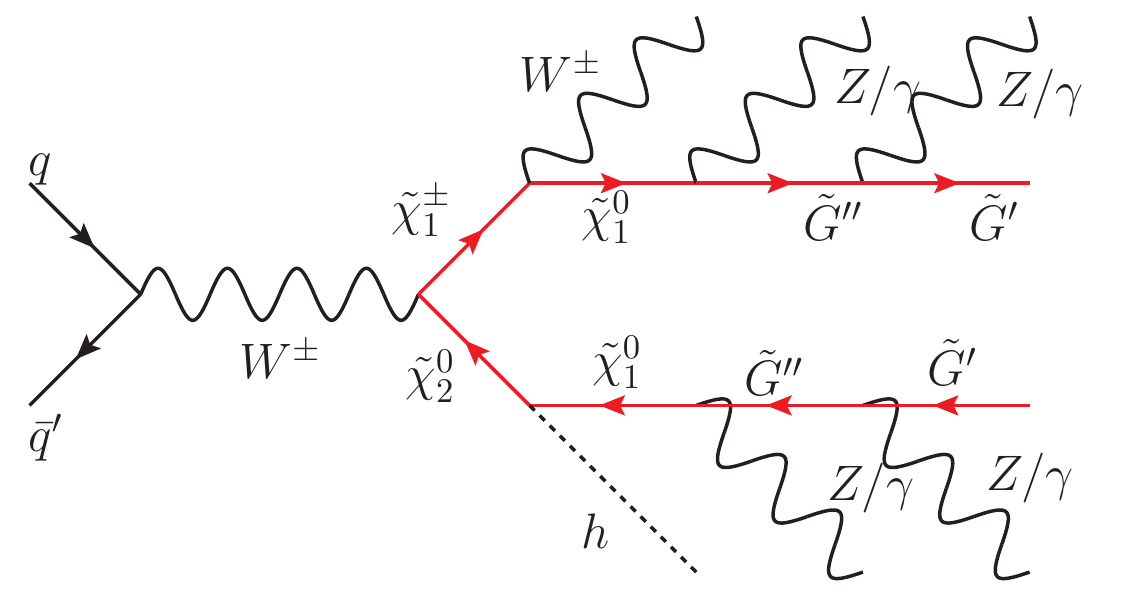}
    \includegraphics[width=0.45\textwidth]{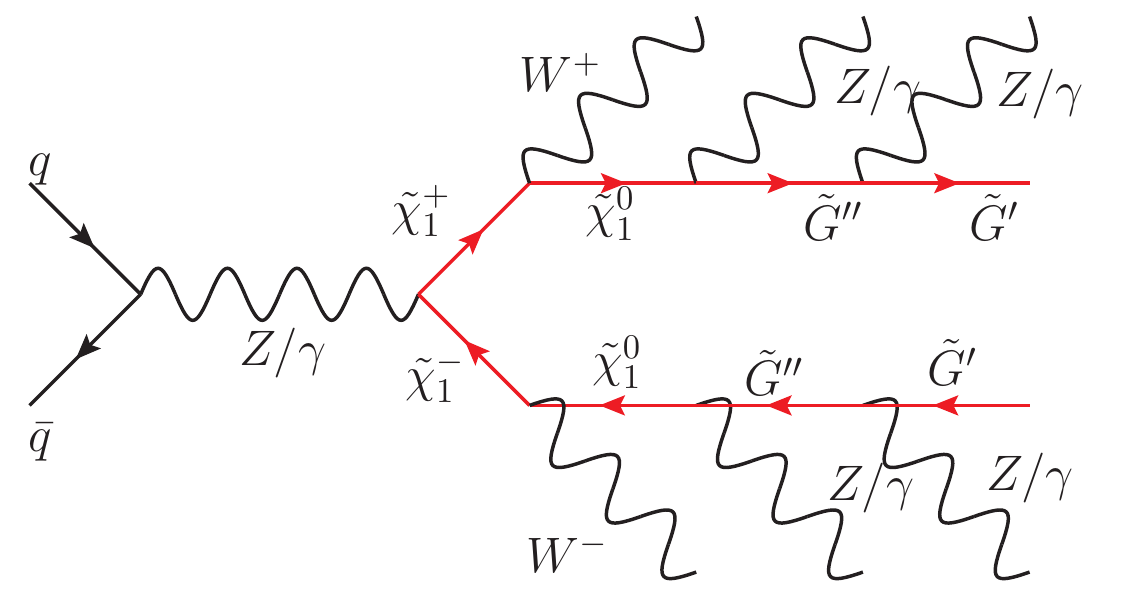}
\caption{Typical signal Feynman diagrams with four $Z/\gamma$. }
    \label{fig:diagrams4a}
\end{figure}

\begin{figure}[htbp]
    \centering
    \includegraphics[width=0.45\textwidth]{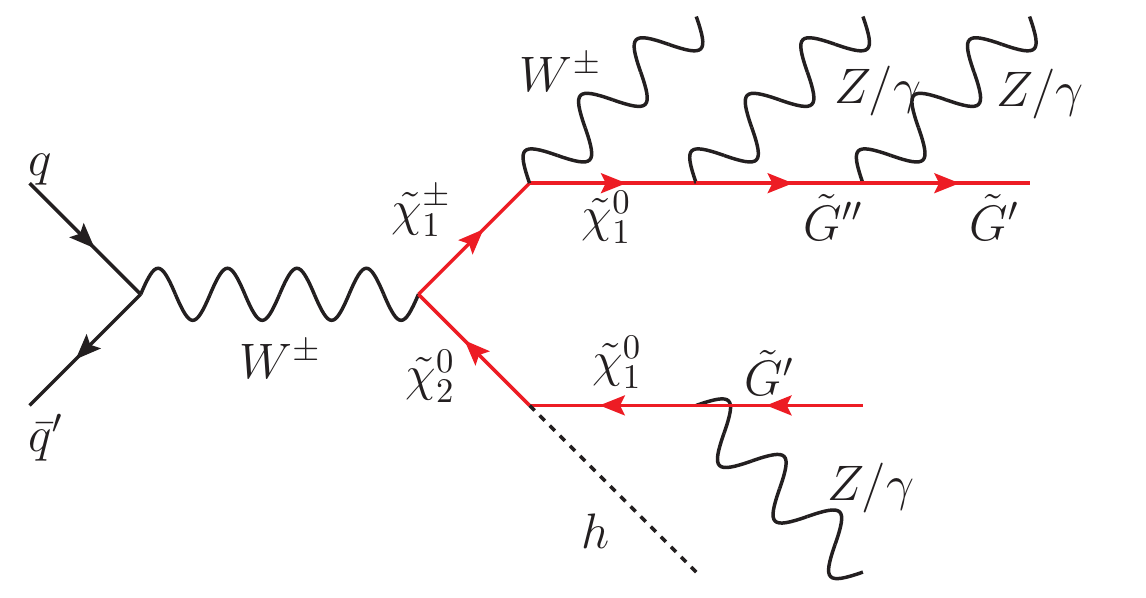}
    \includegraphics[width=0.45\textwidth]{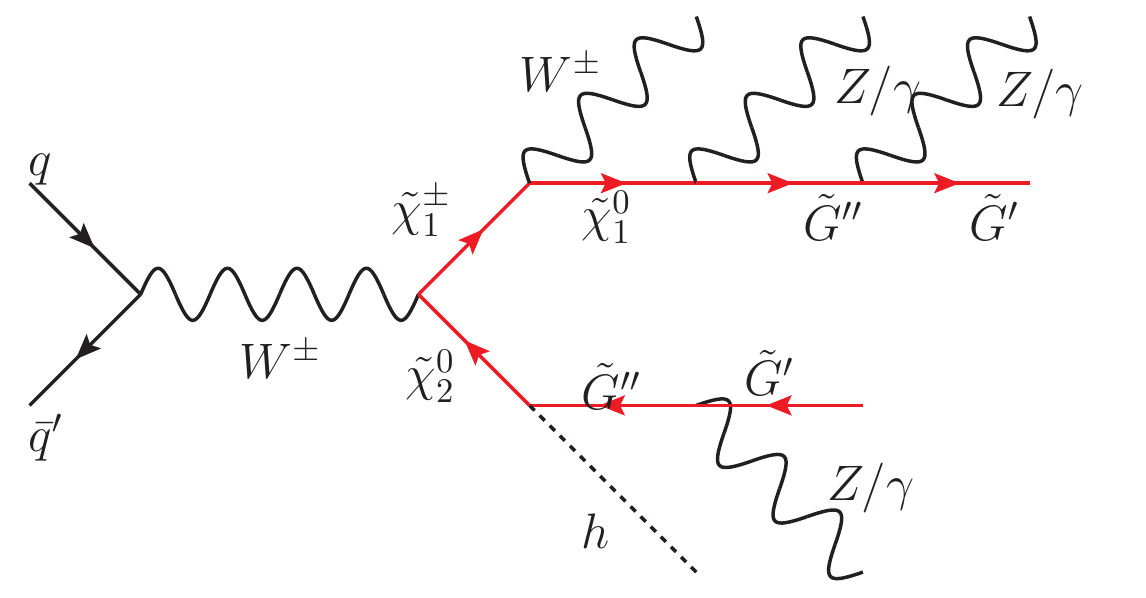}
\caption{Typical signal Feynman diagrams with three $Z/\gamma$. }
    \label{fig:diagrams3a}
\end{figure}

\begin{figure}[htbp]
    \centering
         \includegraphics[width=0.4\textwidth]{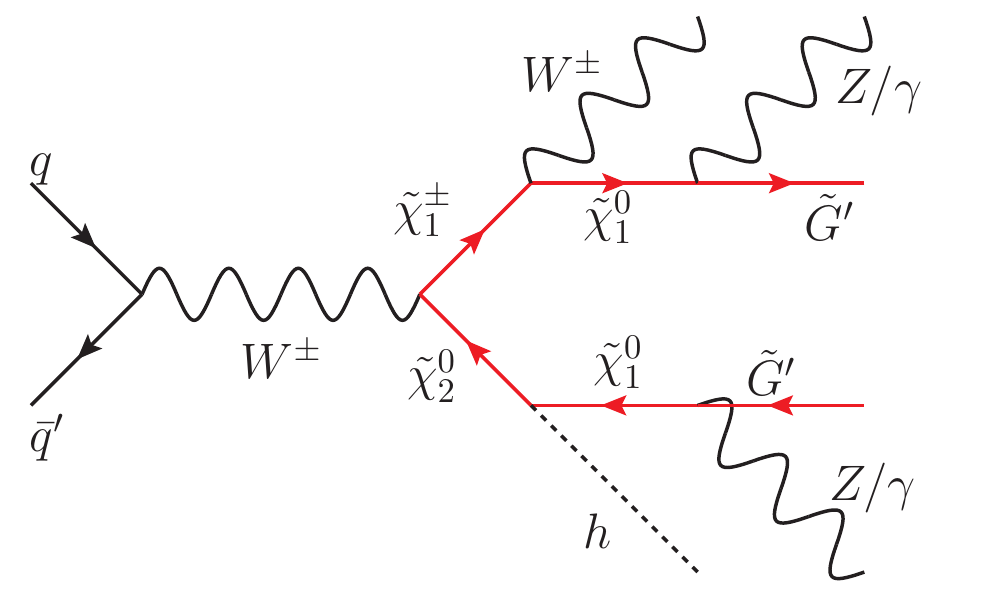}
         \includegraphics[width=0.4\textwidth]{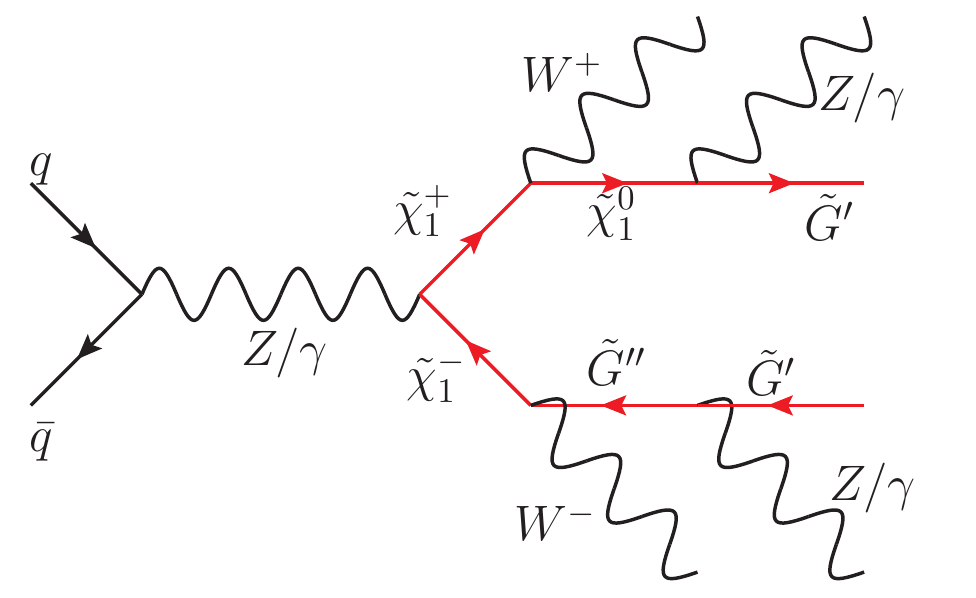}
\caption{Typical signal Feynman diagrams with two $Z/\gamma$. }
    \label{fig:diagrams2a}
\end{figure}

In this configuration the main collider signature will be the production of charginos and neutralinos with their subsequent decay into a cascade of SM particles and lighter neutralinos.
See Figs.~\ref{fig:diagrams4a}-\ref{fig:diagrams2a} for some illustrative leading order (LO) Feynman diagrams.
The decays of the \pgldi are largely determined by the interactions of (\ref{exotic}).
We concentrate on the case where the dominant processes generate three or more photons as depicted in Figs. \ref{fig:diagrams4a} and \ref{fig:diagrams3a}, compared with Figs. \ref{fig:diagrams2a} which has at most two photons in the final state.
We keep $\tilde G'$ in the final state due to the fact that it is typically long-lived. Displaced vertices or long-lived particle searches might be relevant for its discovery but will not be considered here. 
This type of signature is dominant in the case when $m_{\pgldtwo}\lesssim 100\GeV$ and the $M_{W(3)}$ is suppressed so that the wino-like state 
\chinoonepm decays dominantly to \ninoone (and not directly to \pgldthree).

\section{Benchmark points with multi-photon signatures}
\label{sec:benchmarks}

Having constructed the model in Sec.~\ref{sec:model} we can now look for regions of parameter space of the NMSSM with multiple SUSY breaking sectors that
generate signatures with at least three photons. 
Our search for benchmark points proceeds in two steps: first we determine the bosonic sector and then we use the parameters found as input to finding the fermionic sector. 

As far as the bosonic sector is concerned, after solving the tadpole equations, the scalar potential depends on the six parameters $\lambda, k, A_\lambda, A_k, v_s, \tan\beta$. 
Interestingly, it is possible to solve for the soft terms $k$, $A_k$ and $A_\lambda$ in terms of the charged and pseudoscalar masses $m_{H^\pm}$, $m_{a_1}$ and  $m_{a_2}(>m_{a_1})$, with the only extra requirement that $m_{a_2}^2 + m_W^2 > m^2_{H^\pm}$. 
After fixing the values of $m_{H^\pm}$, $m_{a_1}$ and  $m_{a_2}$, one only needs to scan over $\tan\beta$, $v_s$ and $\lambda$ obeying the reality constraint
\beq
    \max(m_{a_1}^2 + m_W^2 - m^2_{H^\pm}, 0) <\lambda^2 < m_{a_2}^2 + m_W^2- m^2_{H^\pm}.
\eeq
Of course, one could also solve for these remaining three parameters as functions of the three scalar masses, but the expressions and the required reality constraints would be quite unwieldy. We believe this is an optimal compromise between analytical and numerical analysis for our purposes.

The scan proceeds as follows. We start by letting the above masses range between $400\mbox{ GeV}\leq m_{a_1}\leq 1000\mbox{ GeV}$, $1000\mbox{ GeV}\leq m_{H^\pm}\leq 1400\mbox{ GeV}$ and $m_{H^\pm}\leq m_{a_2}\leq 1400\mbox{ GeV}$ as well as $1.5\leq \tan\beta\leq 2.5$. 
For each of these choices, we scan over the acceptable range of $\lambda$ and $v_s$, where $\lambda v_s/\sqrt{2} > 1500\mbox{ GeV}$ is required in order to have a heavy higgsino, and check for vacuum stability and a tree-level Higgs mass between 100~GeV and 110~GeV, so that the radiative corrections can raise it to 125~GeV. If we find a solution, we save it and move to the next point in the masses and $\tan\beta$.

For each one of the bosonic benchmark points we then proceed to study the neutralino and chargino sector.
The large parameter space of the fermionic sector ($f_i$, $M_{B(i)}$, $M_{W(i)}$, $m^2_{H_u(i)}$, $m^2_{H_d(i)}$ , $m^2_{S(i)}$, $A_{\lambda(i)}$, $A_{k(i)}$, $\mu_{ij}$) accommodates many different experimental signatures with only a specific subset relevant for multiphoton signatures.
This subset can be defined more precisely by some physical conditions that we describe below.

One requirement is the presence of at least two promptly decaying \pgldi with a partial decay width into photons $\Gamma>10^{-12}\mbox{ GeV}$.
A second requirement is $m_{\pgldthree}<90\GeV$ to avoid decays into $Z$ bosons and $m_{\pgldthree} - m_{\pgldtwo}>30\GeV$ and $m_{\ninoone} -m_{\pgldthree}>30\GeV$ to avoid decays with soft photons.
As a third requirement, a hierarchy in the partial widths into photons is imposed to enhance a decay chain topology
$\Gamma(\ninoone\to\pgldthree\gamma)>\Gamma(\ninoone\to \tilde G^{(')}\gamma)$ and 
$\Gamma(\pgldthree\to \pgldtwo\gamma)>\Gamma(\pgldthree\to \pgldone\gamma)$.
Lastly, we demand the suppression of a direct decay of the lightest chargino into one of the \pgldi, which would also reduce the number of photons in the final state.
After imposing these conditions a total 138 different benchmark points are obtained. These are the targets of our analysis.

The requirement of a prompt decay into photons is what makes it necessary to consider the three hidden sector case instead of the two sector one. 
As had already been hinted by the estimates in~\cite{Ferretti:2013wya}, after a thorough scan we have been unable to find points in the two sector model where the last decay $\pgldtwo \to \pgldone\, \gamma$ is prompt. The signature of the two sector case is thus the same as that of the usual GM scenario but with the massless goldstino replaced by a massive \pgldtwo in the usual decay chains: $\chinoonepm \to W^\pm\,(\ninoone \to \pgldtwo\, \gamma)$ and  $\ninotwo \to Z/h\,(\ninoone \to \pgldtwo\, \gamma)$.
On the contrary, the three sector model can give rise to decay chains containing up to two prompt photons, for a maximum of four photons per process:  $\chinoonepm \to W^\pm\,(\ninoone \to \pgldthree\, \gamma\to \pgldtwo\, \gamma\,\gamma)$ and $\ninotwo \to Z/h\,(\ninoone \to \pgldthree\, \gamma\to \pgldtwo\, \gamma\,\gamma)$.

\section{Simulation of multi-photon processes}
\label{sec:simulation}

The sensitivity of the ATLAS experiment to the NMSSM+GM models with multi-photon final states is estimated with simulated signal and background events in $pp$ collisions at $\sqrt{s}=13\TeV$ center-of-mass energy. Events are generated with the \mg program~\cite{Alwall:2011uj} using default dynamical factorization and renormalization scales and the NNPDF 3.1 NLO LUXQED parton distribution function (PDF) set with $\alpha_s(\mu)=0.118$~\cite{Ball:2013hta}. 
Parton level events are processed through \pythia~\cite{Sjostrand:2014zea} for showering and hadronization and through \delphes 3~\cite{deFavereau:2013fsa} for fast detector simulation. The default \pythia and ATLAS \delphes~cards are used.

Guided by the mass ranges populated by the benchmark points described in the previous section, signal events are generated in a grid scanning  ($m_{\chinoonepm}=m_{\ninotwo},\, m_{\ninoone}$)
with  $m_{\pgldthree}=80\GeV$ and $m_{\pgldtwo}=50\GeV$.
Besides this so-called chargino-neutralino grid, signal events are also generated in the following complementary grids: ($m_{\ninoone},\, m_{\pgldthree}$) with $m_{\chinoonepm}=m_{\ninotwo}=1.4\TeV,\,m_{\pgldtwo}=50\GeV$, ($m_{\pgldthree},\, m_{\pgldtwo}$) with $m_{\chinoonepm}=m_{\ninotwo}=1.2\TeV,\,m_{\ninoone}=300\GeV$ and
($m_{\chinoonepm}=m_{\ninotwo},\,m_{\ninoone}$) with $m_{\pgldthree} = (m_{\ninoone} + m_{\pgldtwo})/2,\,m_{\pgldtwo}=50\GeV$.

The signal generation is restricted to the typically dominating process
\begin{equation}
pp\to (\chinoonepm \to W^\pm\,(\ninoone \to \pgldthree\, \gamma\to \pgldtwo\, \gamma\,\gamma)) 
    + (\ninotwo \to h\,(\ninoone \to \pgldthree\, \gamma\to \pgldtwo\, \gamma\,\gamma))
\label{eq:dominantproc}
\end{equation}
depicted in the left diagram of \fig{fig:diagrams4a}, with only the decay chains involving photons considered. All other processes in \figs{fig:diagrams4a}{fig:diagrams2a} are generated only for a limited set of masses to allow for comparisons of the selection efficiencies. The event generation and total cross section calculation are performed at leading order (LO) with the aid of a \textsc{Universal Feynrules Output} (UFO)~\cite{Degrande:2011ua} model implemented locally
\footnote{The \textsc{FeynRules} and UFO model can be retrieved from the High Energy Model Database \url{https://hepmdb.soton.ac.uk/hepmdb:1121.0333}.} 
through the \textsc{FeynRules} package~\cite{Alloul:2013bka, Christensen:2008py} and based on the NMSSM implementation~\cite{Conte:2016zjp}. The model includes the SM tree-level interactions and the interactions described in \sec{sec:model}. The decay chain syntax, which adopts the spin correlated narrow width approximation~\cite{Frixione:2007zp}, is used.

The simulation of the SM background is carried out analogously. The following backgrounds are simulated at LO: $pp\to Z/W^\pm+\gamma\gamma$, $pp\to Z/ W^\pm+\gamma\gamma\gamma$, $pp\to \gamma\gamma$, $pp\to \gamma\gamma\gamma$ and $pp\to t\bar{t}+\gamma\gamma$. 
Based on these simulations, the dominant backgrounds after applying the selection criteria defined in Eq.~\ref{eq:selection}, are found to be $pp\to W^\pm+\gamma\gamma$ and 
$pp\to \gamma\gamma$, where the third identified photon is {\it fake}, i.e. it originates from hadronic activity (mostly pions decaying into photons) or from misreconstructed electrons.  
For these processes, simulations at next-to-leading order (NLO) in QCD are performed and used in the analysis. All simulations are carried out with the parton-level cuts reported in \tab{tab:parton_cuts}.

To avoid double counting we also perform a matching for samples with different photon multiplicities according to the prescription introduced in \cite{BuarqueFranzosi:2020baz}. 
The observed effect of the matching is found to be negligible.

\begin{table}[htbp]
\centering
\begin{tabular}{|c|c|c|}
\hline
$p_T(j)>20\GeV$ & $\Delta R(j,\gamma)>0.4$  & $|\eta(\gamma)|<2.5$ \\
$p_T(\ell)>10\GeV$ & $\Delta R(\ell^+,\ell^-)>0.4$ & \\
$p_T(\gamma)>10\GeV$ & $\Delta R(\ell,\gamma)>0.4$  & \\
\hline
\end{tabular}
\caption{Parton level cuts used in the generation of events for all the background samples, when they are applicable.}
\label{tab:parton_cuts}
\end{table}

\section{Object definition and event selection}
\label{sec:selection}

The signal events of interest are characterized by at least three photons, large missing transverse momentum from the presence of the two long-lived \pgldtwo, and a large amount of activity from the visible decay products of the charginos and neutralinos. The objects used in the analysis are defined in the default ATLAS \delphes\ card~\cite{deFavereau:2013fsa}. Here we give a brief overview of their main characteristics.

\textbf{Photons}, $\gamma$, are reconstructed from energy deposits in the electromagnetic calorimeter (ECAL) with no matching track in the inner tracking system. 
Photons are required to have a $\pT > 30\,\GeV$, $ |\eta|  < 2.37$ and an isolation $I<0.12$ within the cone $\Delta R<0.5$. 
The isolation variable is defined by the scalar sum of the \pT of all objects (not including the candidate) within a cone of $\Delta R$ around the candidate divided by the candidate $p_T$, 
\begin{equation}
I = \frac{\sum_{i\neq \text{candidate}} \pT(i)}{\pT(\text{candidate})}.
\end{equation}

\textbf{Leptons}, $\ell$, refers to  {electrons} and {muons} only, not $\tau$ leptons.
Electrons are reconstructed from energy deposits in the ECAL and a matching track in the inner tracking system. 
Muons are reconstructed from hits in the inner tracking system and the muon spectrometer. 
Leptons are required to have $\pT > 25\, \GeV$ and $| \eta | < 2.47$, and to be isolated with $I<0.12$ and $I<0.25$ for electrons and muons respectively, both within the cone $\Delta R<0.5$.

\textbf{Jets}, $j$, are reconstructed with the \fastjet~\cite{Cacciari:2011ma} package using the anti-$k_t$ algorithm~\cite{Cacciari:2008gp} with a distance parameter $R=0.4$.
They are required to have $\pT > 25\GeV$ and $| \eta | < 2.47$.

\textbf{Missing transverse energy}, \met, is calculated as the negative vector sum of the transverse momenta of all reconstructed basic objects: photons, electrons, muons and jets: $\vec{E}_{\text{T}}^{\text{miss}} = - \sum_i \vec{\pT}(i)$~\cite{deFavereau:2013fsa}. 

\textbf{Scalar transverse energy}, \HT, is calculated as the scalar sum of the transverse momenta of the same basic objects as those used in the \met calculation. 

By optimizing the discovery significance of the chargino-neutralino grid point with
($m_{\ninotwo}=m_{\chinoonepm}[\text{GeV}]$,$\, m_{\ninoone}[\text{GeV}]$) = ($1000,780$),
while keeping at least 80 MC events for the diphoton background, the following three-photon signal region \SRggg was obtained:
\begin{equation}
n_\gamma\geq 3,\quad \met>50\GeV,\quad \HT >400\GeV\,
\label{eq:selection}
\end{equation}
We note that the choice of selection criteria is partially driven by limited MC statistics and is thus not fully optimal.  In particular a harder cut on $\HT$ is expected to further improve the sensitivity.

In addition to the three-photon signal region \SRggg, we also investigate the sensitivity of the two-photon signal region \SRWH (referred to in this paper as \SRgg) from \cite{Aaboud:2018doq}. This signal region requires 
\begin{equation}
\begin{aligned}
    n_\gamma\geq 2\ (E_T^\gamma>75\GeV),\quad\met > & 250\GeV, \quad H_T>1000\GeV, \\ \Delta\phi_{\text{min}}(j,\met)>0.5, \quad & \Delta\phi_{\text{min}}(\gamma,\met)>0.5
\end{aligned}
\label{eq:SRgg}
\end{equation}

\section{Background estimates}
\label{sec:background_estimates}

The rate of fake photons in the $W^\pm+\gamma\gamma$ sample is substantial, originating from a misreconstructed electron from the $W^\pm\to e^\pm \nu$ decay or from the $W^\pm\to \tau^\pm \nu\to e^\pm \nu\nu$ cascade decay, and is parametrized by a table of efficiencies in the \delphes~card.
On the other hand, the rate of fake photons in the $pp\to \gamma\gamma$ sample is significantly lower. Here, fake photons mainly arise from hadronic activity, which is quite suppressed by requiring the photons to have large \pT and to be isolated from other activity in the calorimeters and tracking detectors. In order to avoid having to simulate a prohibitive amount of events we adopt a parametrization of fake photons originating from jets, i.e. we select a jet instead of a photon and apply a reweighting of the event according to the fake rate~\cite{ATL-PHYS-PUB-2013-009}:
\begin{equation}
f(\pT)=9.3\times 10^{-3}\exp(-\pT/27.5\GeV)\,.
\label{eq:fakephoton}
\end{equation}

The next most relevant background is $t\bar{t}+\gamma\gamma$ (in the fully leptonic decay mode) and $W^\pm+\gamma\gamma\gamma$, each contributing with less than 5\% of the $W^\pm+\gamma\gamma$ sample.

For the systematic uncertainties related to the modeling of fake photons, estimates are based on those found in the ATLAS search~\cite{Aaboud:2018doq}. For the $\gamma\gamma$ process, where fake photons are modeled by Eq.~\ref{eq:fakephoton}, we use a $\pm 100\%$ uncertainty, while a $\pm 30\%$ uncertainty is used for the $W^\pm+\gamma\gamma$ samples (and others including $Z+\gamma\gamma$ and $t\bar{t}+\gamma\gamma$)  where the modeling of fake photons from electron and hadronic activities is done by \delphes.

The systematic uncertainty on the total cross sections of these backgrounds associated with the choice of scales is assessed by considering seven different values for the renormalization and factorization scales. Based on an envelope of the seven samples, the uncertainties on the $\gamma\gamma$ and $W^\pm+\gamma\gamma$ cross sections are estimated to be $\pm 20\%$ and $\pm 7\%$ respectively.
Since radiative corrections at NNLO are large at high values of \HT~\cite{Gehrmann-DeRidder:2018kng}, an uncertainty of $\pm 50\%$ is assumed for the $W^\pm+\gamma\gamma$ process in the \SRggg even though our computation at NLO gives only a $\pm 7\%$ scale dependence on the total cross section. 
For the $\gamma\gamma$ sample we use the NLO scale dependence of $\pm 20\%$. 
For the other processes we consider a 60\% uncertainty since they were estimated at LO. 

The uncertainties on the background estimates are added in quadrature and are summarized in \tab{tab:systematics}. The backgrounds included in "others" are $t\bar{t}+\gamma\gamma$, $W^\pm+\gamma\gamma\gamma$, $Z+\gamma\gamma$ and $Z+\gamma\gamma\gamma$ (the last two in the $Z\to \ell^+\ell^-,\nu\bar{\nu}$ decay mode).

\begin{table}[htbp]
\centering
\begin{tabular}{|c|c|c|c|}
\hline
										& Fake photon &  Radiative corrections & Total \\
\hline
$W^\pm + \gamma\gamma$					&  30\%	& 50\%  & 58\%\\
$\gamma\gamma$							&  100\% & 20\% & 102\%\\
others									&  30\% & 60\%  & 67\%\\
\hline
\end{tabular}
\caption{Systematic uncertainties on the background estimates.}
\label{tab:systematics}
\end{table}

The effective cross sections of the background processes after parton-level cuts (see \tab{tab:parton_cuts}), $\sigma_0$, and after the \SRggg selection criteria (see Eq. \ref{eq:selection}), $\sigma_{\SRggg}$, are shown in \tab{tab:bg_samples}, including systematic uncertainties for the latter.
For a given integrated luminosity $\mathcal{L}$ in fb$^{-1}$, this results in an expected number of background events
\begin{equation}
B = (0.0399\pm 0.0270) \times \mathcal{L}\,[\ifb],
\end{equation}
where only systematic uncertainties are included.

\begin{table}[htbp]
\centering
\begin{tabular}{|c|c|c|c|c|}
\hline
										& $\sigma_0$ [fb]				& $\sigma_{\SRggg}$ [fb] \\
\hline
$W^\pm + \gamma\gamma$						&  $1.87\times 10^{2}$ 	& $0.029 \pm 0.017$\\
$\gamma\gamma$							&  $1.93\times 10^{5}$ & $0.00811 \pm 0.00827$ \\
others									&  $1.28\times 10^{5}$ & 
$0.00278 \pm 0.00186$ \\
\hline
Total & $3.21 \times 10^5 $ & $0.0399 \pm 0.0270$\\
\hline
\end{tabular}
\caption{Cross sections (in fb) after parton-level cuts (\tab{tab:parton_cuts}) only, $\sigma_0$, and after the \SRggg selection criteria (Eq.~\ref{eq:selection}), $\sigma_{\SRggg}$.}
\label{tab:bg_samples}
\end{table}

\section{Results}
\label{sec:results}

We perform a typical cut and count analysis in \SRggg and \SRgg and use the following Asimov formula to extract bounds on the signal cross section~\cite{Li:1983fv, Cousins:2007yta, Cowan:2010js}:
\begin{equation}
z=\sqrt{2 \left((B+S) \log \left(\frac{\left((B \delta )^2+B\right) (B+S)}{B^2+(B \delta )^2 (B+S)}\right)-\frac{B^2 \log \left(\frac{S (B \delta )^2}{B \left((B \delta )^2+B\right)}+1\right)}{(B \delta )^2}\right)}\,.
\label{eq:asimov}
\end{equation}
Here, $S$ and $B$ are the estimated number of signal and background events respectively, and $\delta$ is the relative systematic uncertainty on the background estimate. 
The expected upper limit at 95\% confidence level on the number of signal events, $S_{\text{exp}}^{95}$, is defined as the value of $S$ which gives $z=2$. From $S_{\text{exp}}^{95}$, upper limits on the effective signal cross section $\sigma_{\text{exp}}^{95}=S_{\text{exp}}^{95}/\mathcal{L}$ are
derived. Limits on the production cross section of the signal are derived by dividing $\sigma_{\text{exp}}^{95}$ by the selection efficiency, either $\epsilon_{3\gamma}$ or $\epsilon_{\gamma\gamma}$, of a given signal model.

Four luminosity scenarios are considered: LHC Run 1 with $\mathcal{L}=36.1\ifb$\footnote{To be compared with the ATLAS search~\cite{Aaboud:2018doq}.}, LHC Run 2 with $\mathcal{L}=139\ifb$, LHC Run 3 with $\mathcal{L}=300\ifb$ and HL-LHC with $\mathcal{L}=3000\ifb$.
The expected upper limits from \SRggg and \SRgg on the number of signal events, $S_{\text{exp}}^{95}$, and on the effective signal cross section, $\sigma_{\text{exp}}^{95}$, are given for each luminosity scenario in \tab{tab:bounds}.

The upper limits on the production cross section times branching ratio of the main process (Eq.~\ref{eq:dominantproc}) are shown in \fig{fig:bounds_grid} for the chargino-neutralino grid.
To extract these bounds, the selection efficiencies $\epsilon_{3\gamma}$ and $\epsilon_{\gamma\gamma}$ are derived and used for each grid point. The $\epsilon_{\gamma\gamma}$ computation is validated by a comparison with the ATLAS analysis~\cite{Aaboud:2018doq} (see Appendix~\ref{sec:comparison} for details). 
As can be seen, \SRggg gives more stringent limits than \SRgg for the main process throughout the entire mass grid. 
The improvement in sensitivity is particularly significant in the low mass region reaching between one and two orders of magnitude.
Therefore, if final states with three or more photons are as common as those with less than three, our strategy would ensure a better sensitivity. This is the case for the benchmark points discussed in \sec{sec:benchmarks}.

The upper limits on the production cross section times branching ratio for the complementary grids are shown in Fig.~\ref{fig:bounds_SRWH}. Also here, the \SRggg performs better in the whole mass region considered, and is particularly effective in regions of low masses and small mass splitting.

\begin{table}[htbp]
    \centering
    \begin{tabular}{c|c|c|c|c}
    & \multicolumn{2}{c|}{\SRggg} & \multicolumn{2}{c}{\SRgg} \\
    \hline
         $\mathcal{L} [\ifb]$ & $S_{\text{exp}}^{95}$ & $\sigma_{\text{exp}}^{95}$ [fb]  & $S_{\text{exp}}^{95}$ & $\sigma_{\text{exp}}^{95}$ [fb]  \\
         \hline
        36.1     & 4.74       & 0.131  &  3.7*      & 0.103* \\
         
         139      & 13.3       & 0.0957  &  9.31      & 0.0669 \\
         300      & 26.4       & 0.0880  &  16.7      & 0.0556 \\
         3000     & 245.        & 0.0816  &  135.      & 0.0453 \\
    \end{tabular}
    \caption{Upper limits at 95\% C.L. on the number of signal events and the corresponding upper limits on the signal cross section for \SRggg and \SRgg. *Observed limit obtained from~\cite{Aaboud:2018doq}.}
    \label{tab:bounds}
\end{table}

\begin{figure}[phtb]
\centering
\includegraphics[width=0.49\textwidth]{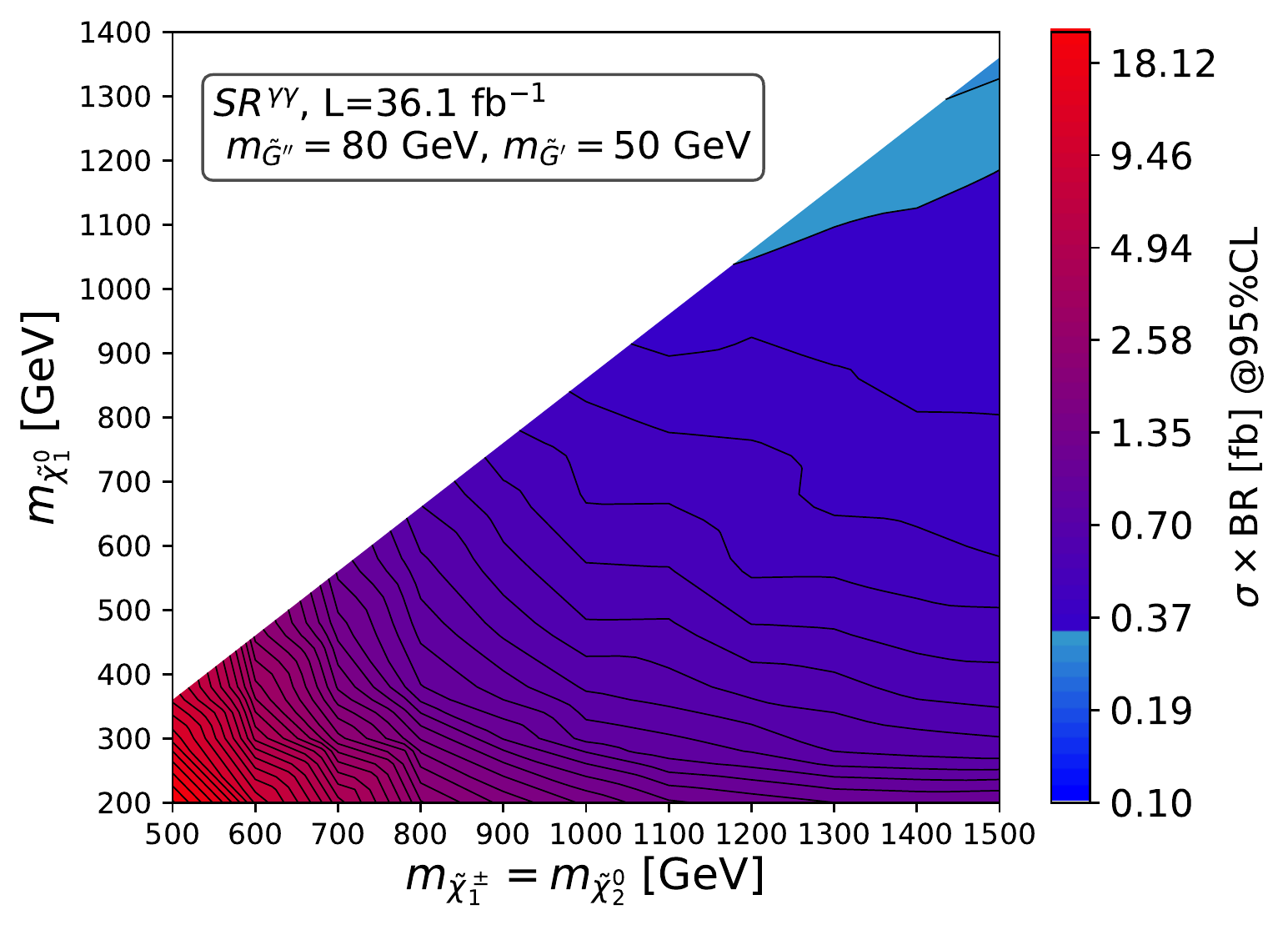}
\includegraphics[width=0.49\textwidth]{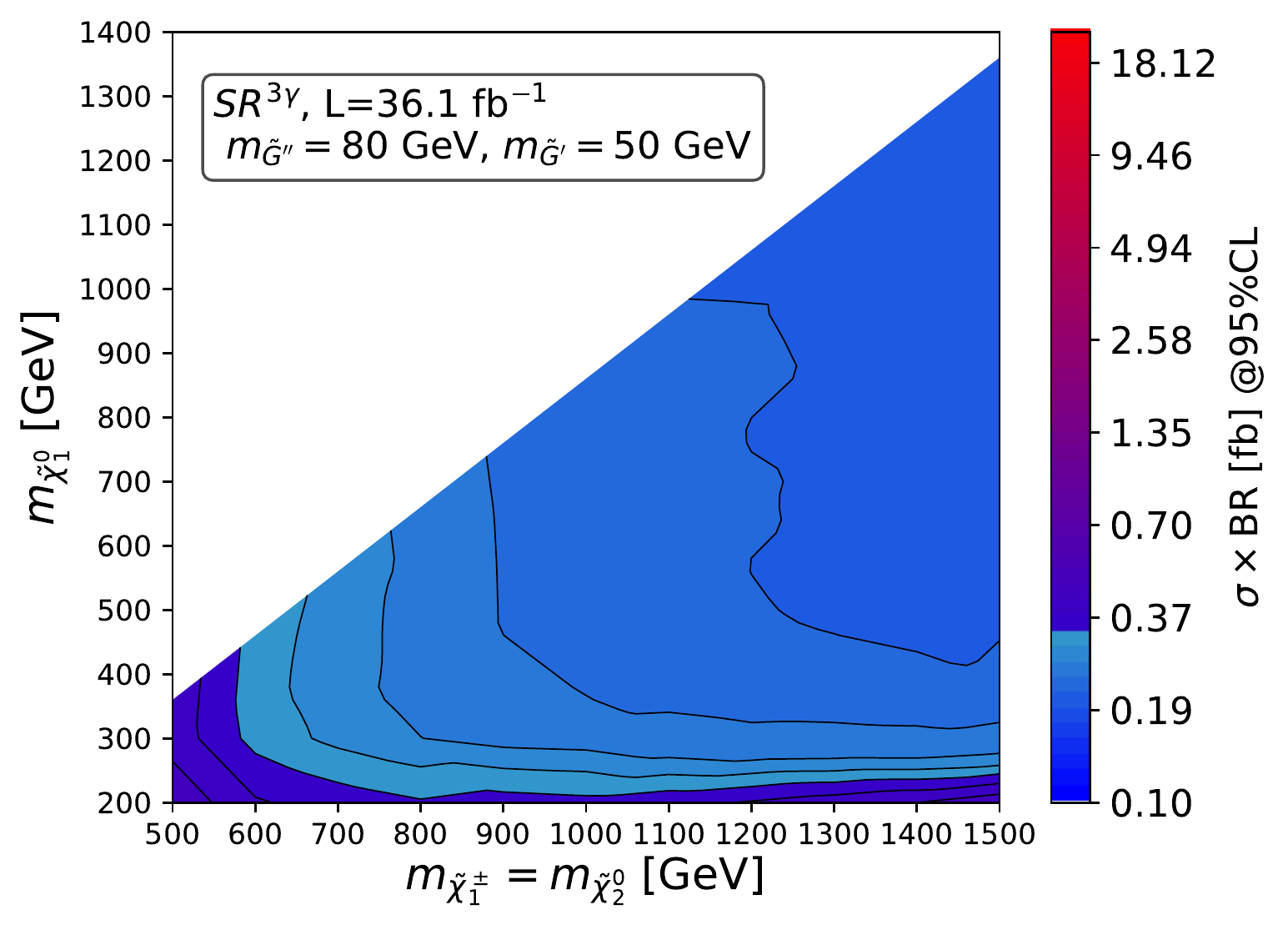}
\includegraphics[width=0.49\textwidth]{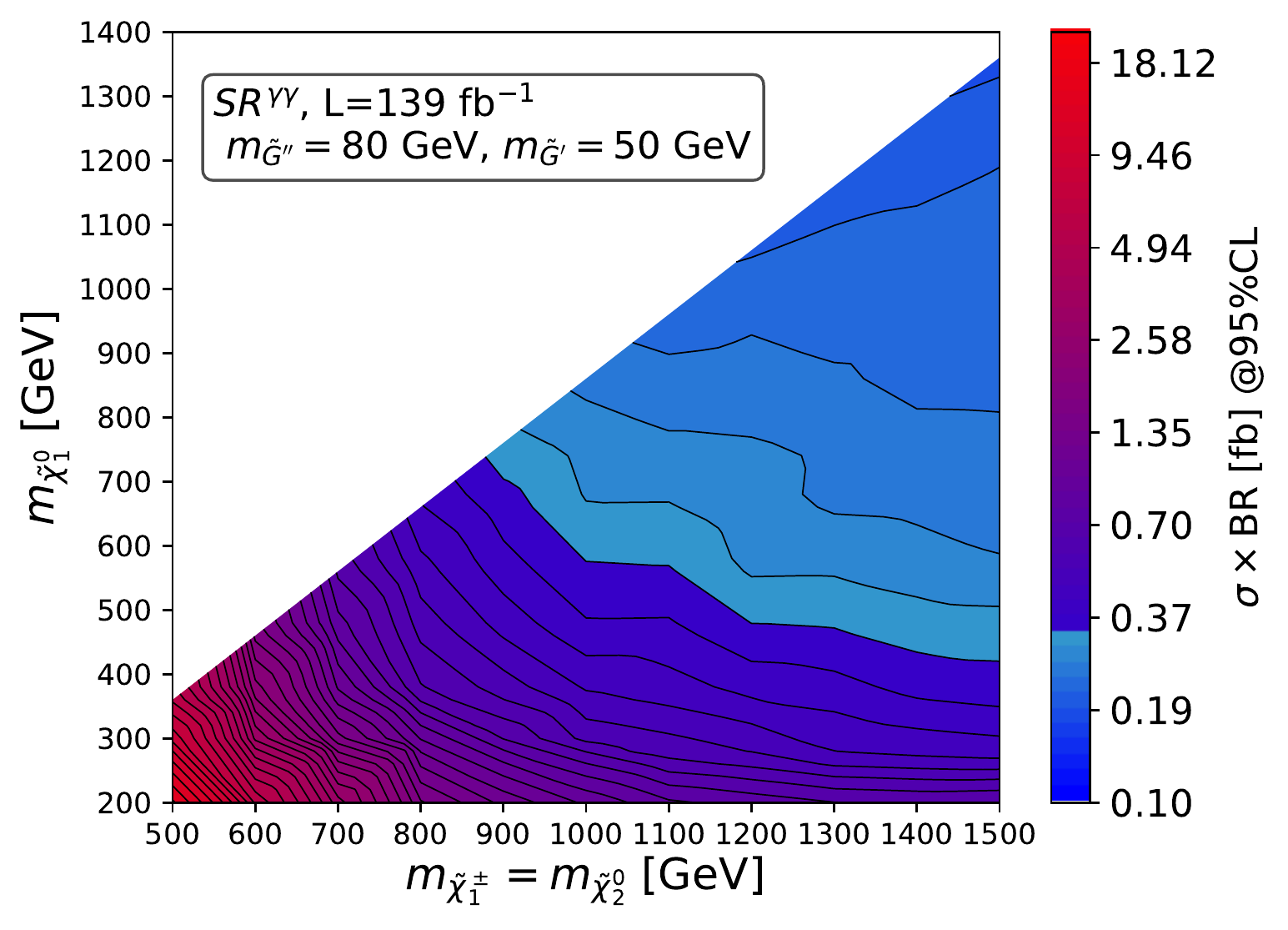}
\includegraphics[width=0.49\textwidth]{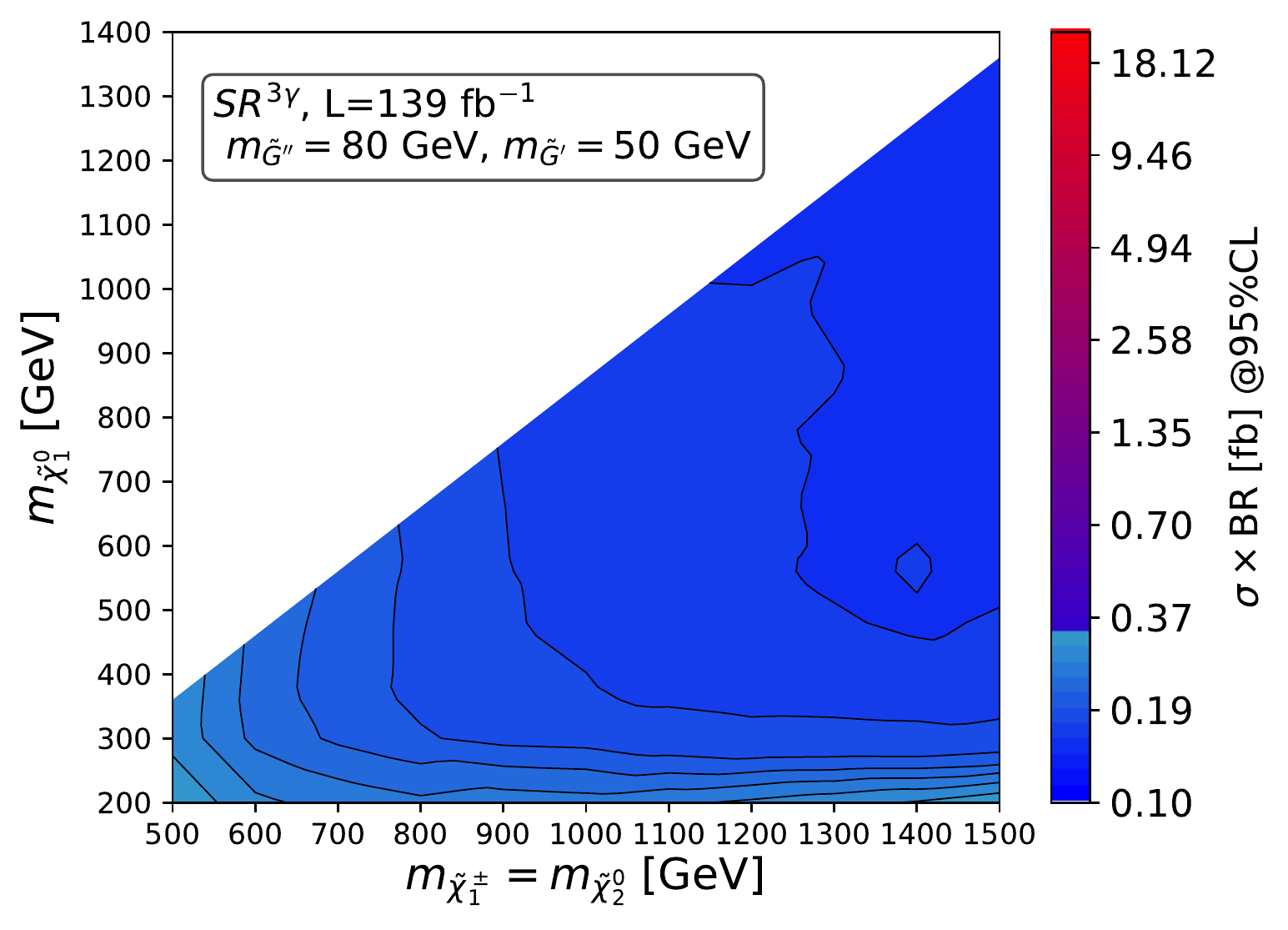}
\includegraphics[width=0.49\textwidth]{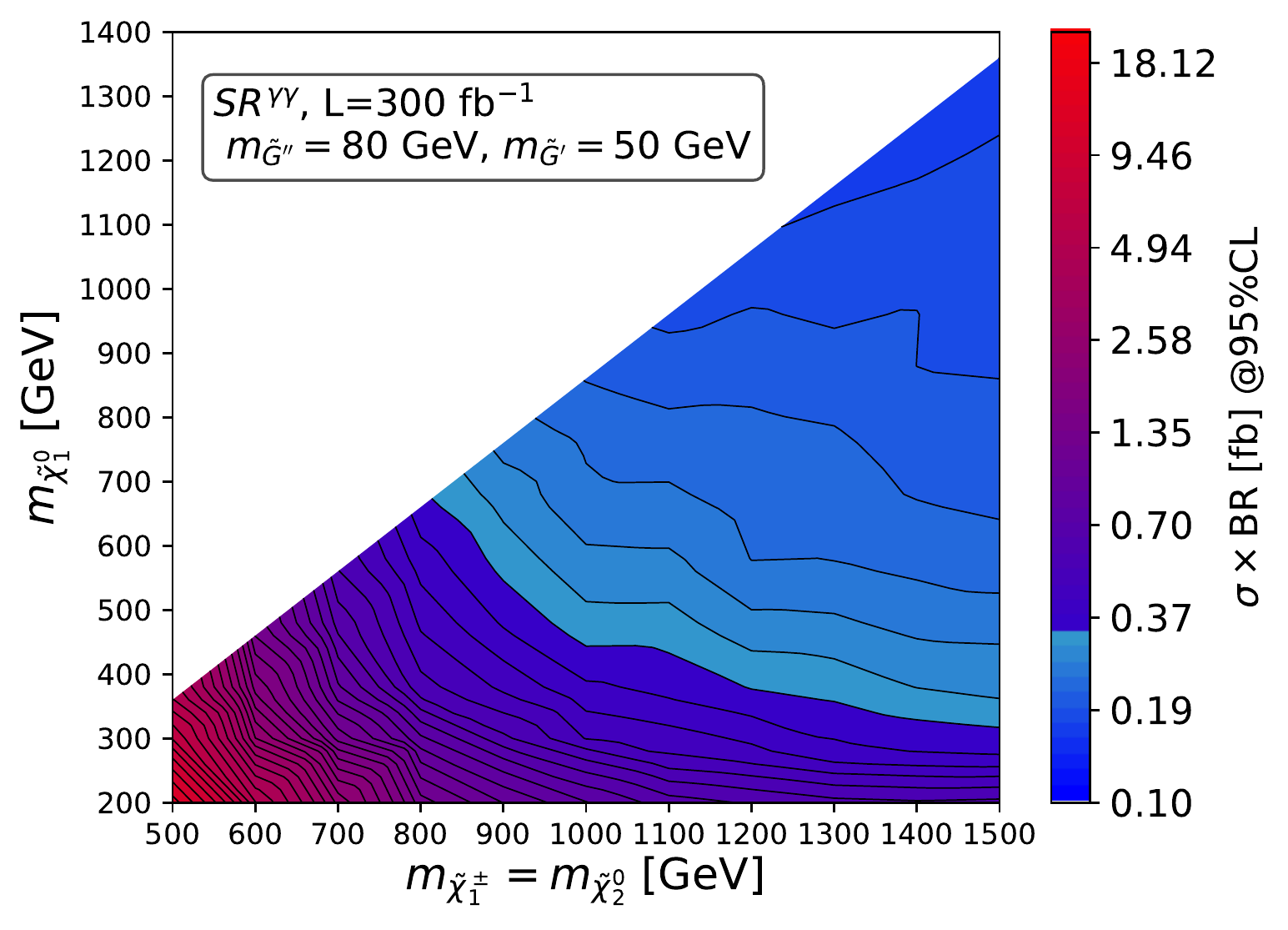}
\includegraphics[width=0.49\textwidth]{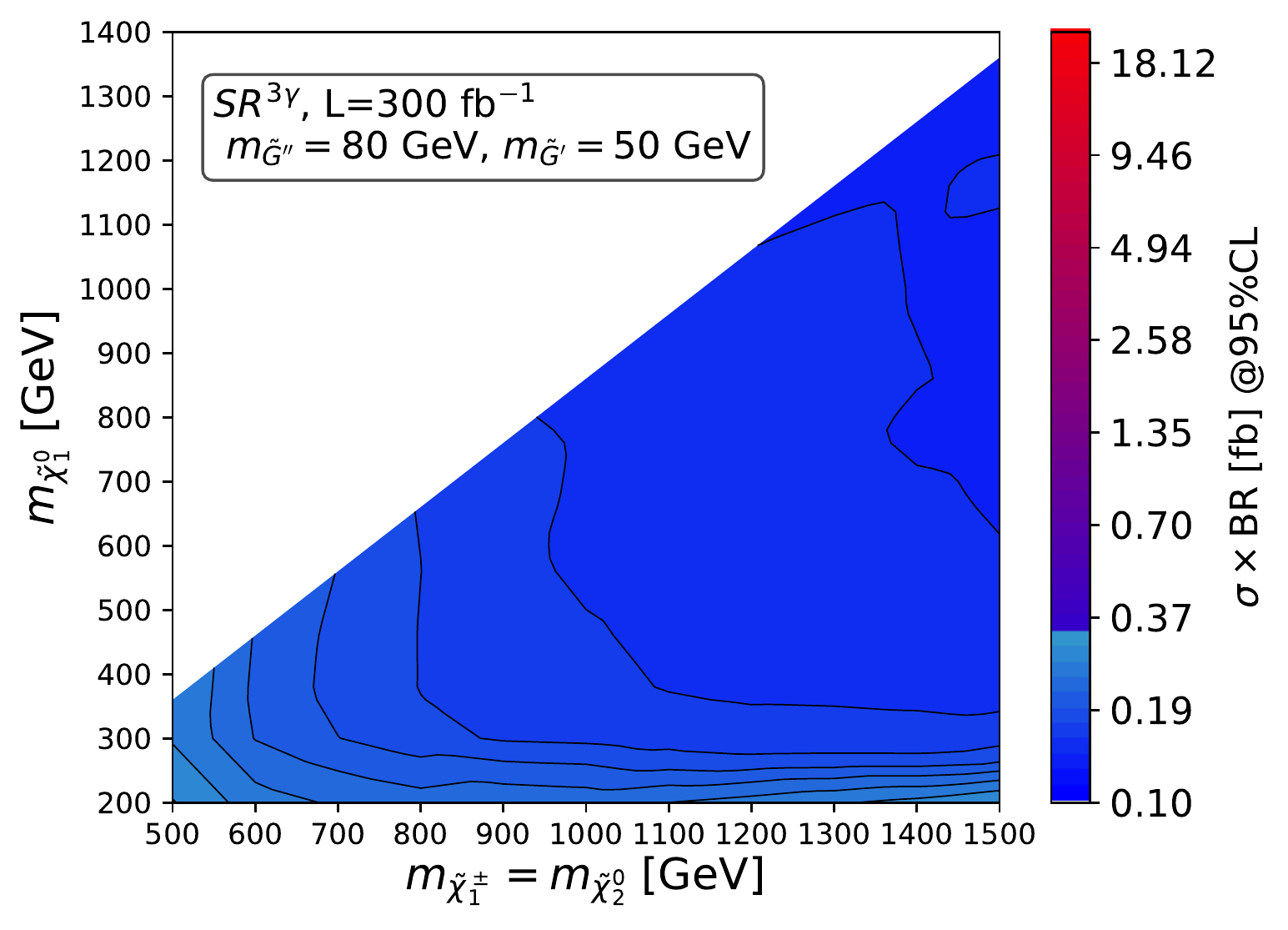}
\includegraphics[width=0.49\textwidth]{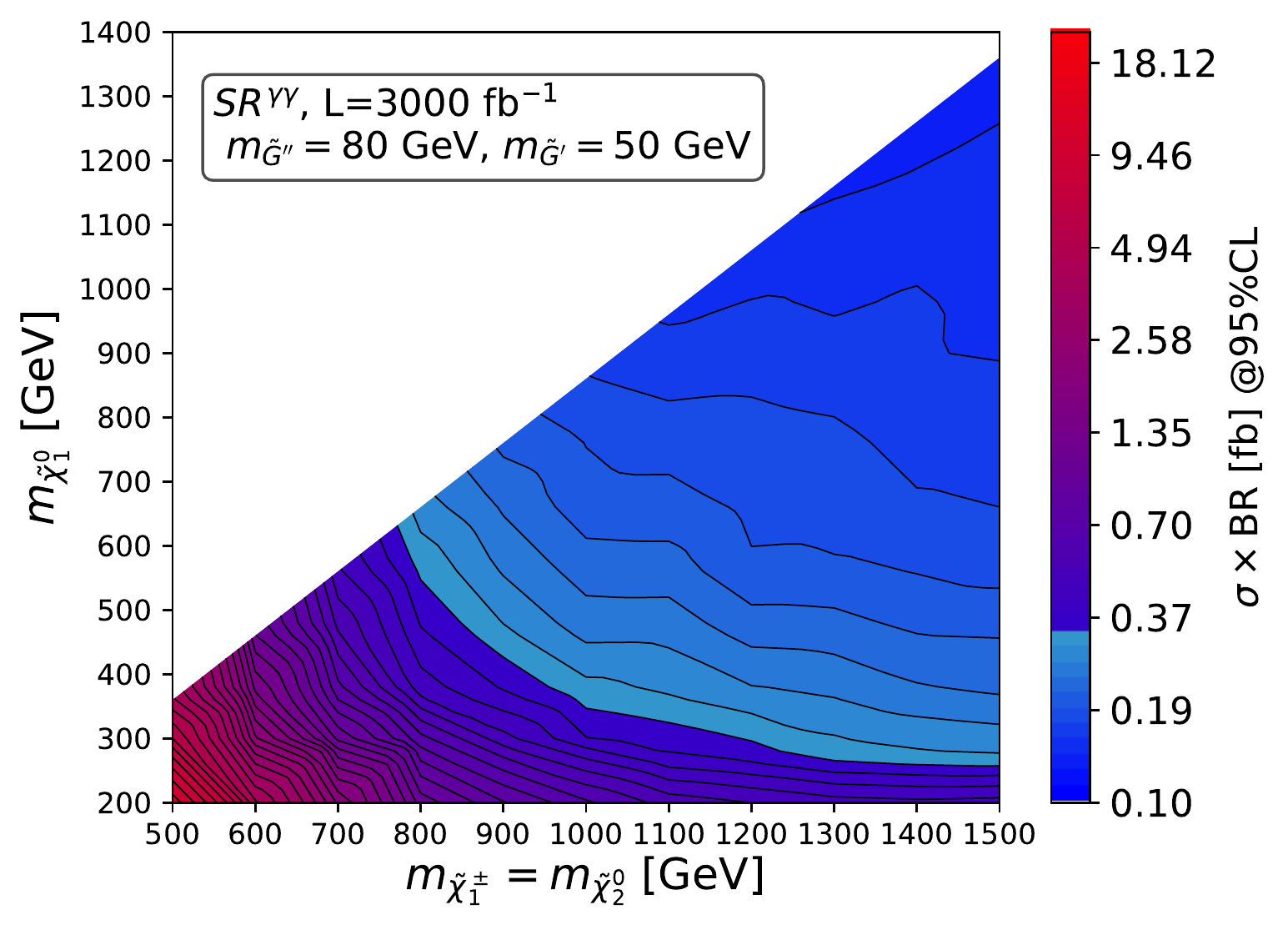}
\includegraphics[width=0.49\textwidth]{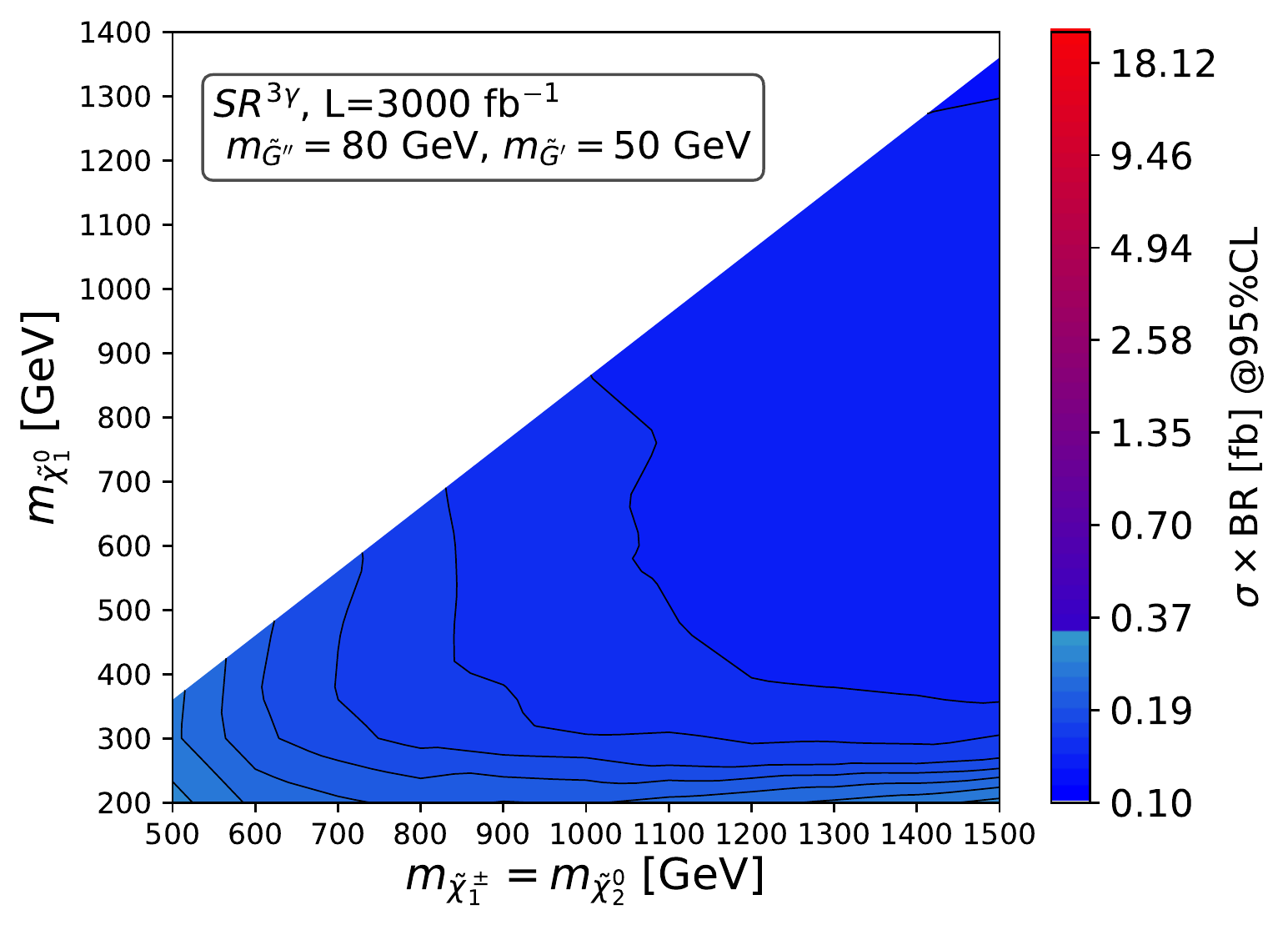}
\caption{The upper limit at 95\% C.L. on the production cross section times branching ratio (in fb) in the chargino-neutralino grid from \SRgg (left) and \SRggg (right). The rows correspond to integrated luminosities of $\mathcal{L}=36.1\ifb$, $139\ifb$, $300\ifb$ and $3000\ifb$.}
\label{fig:bounds_grid}
\end{figure}

\begin{figure}
\centering
\includegraphics[width=0.45\textwidth]{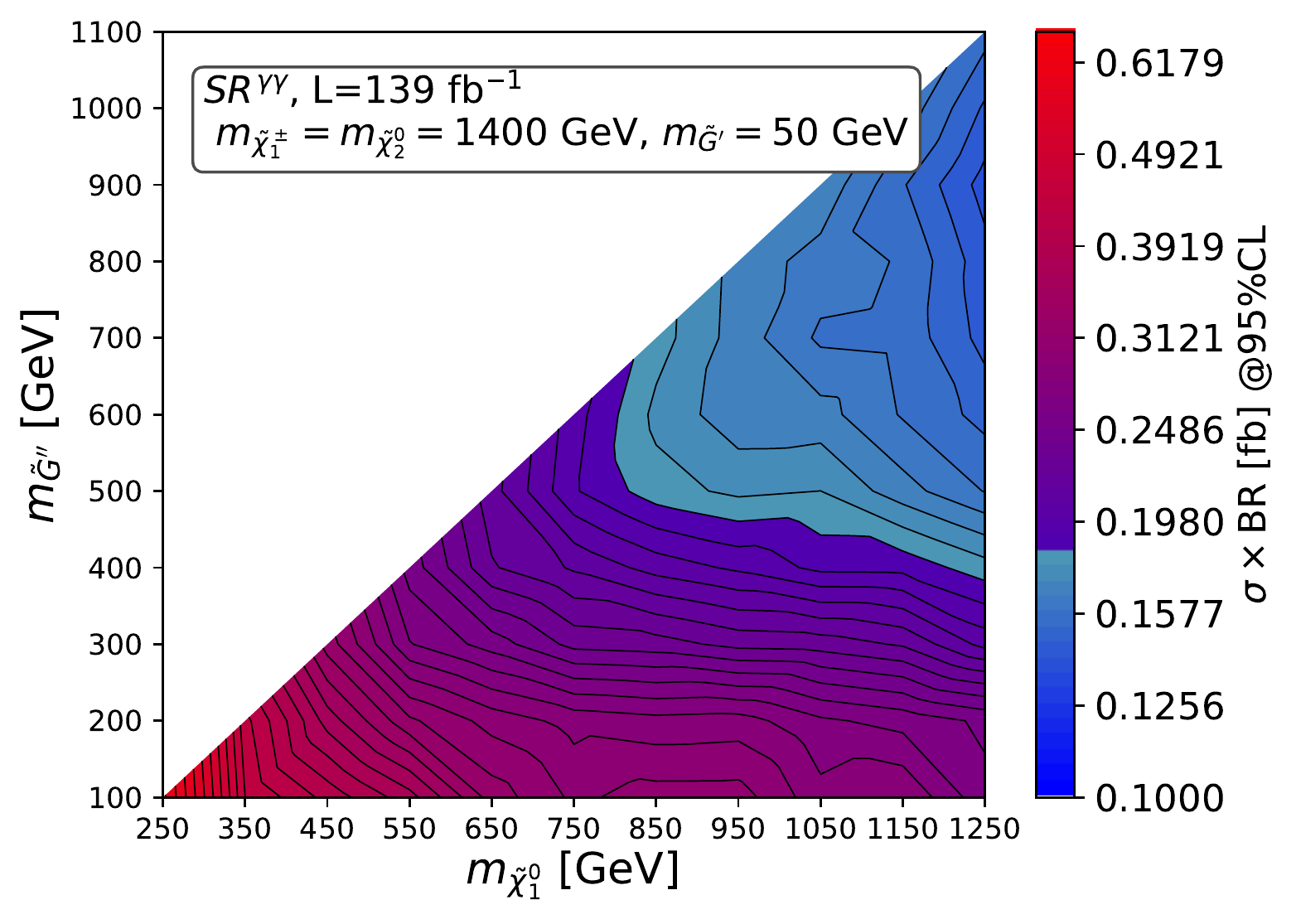}
\includegraphics[width=0.45\textwidth]{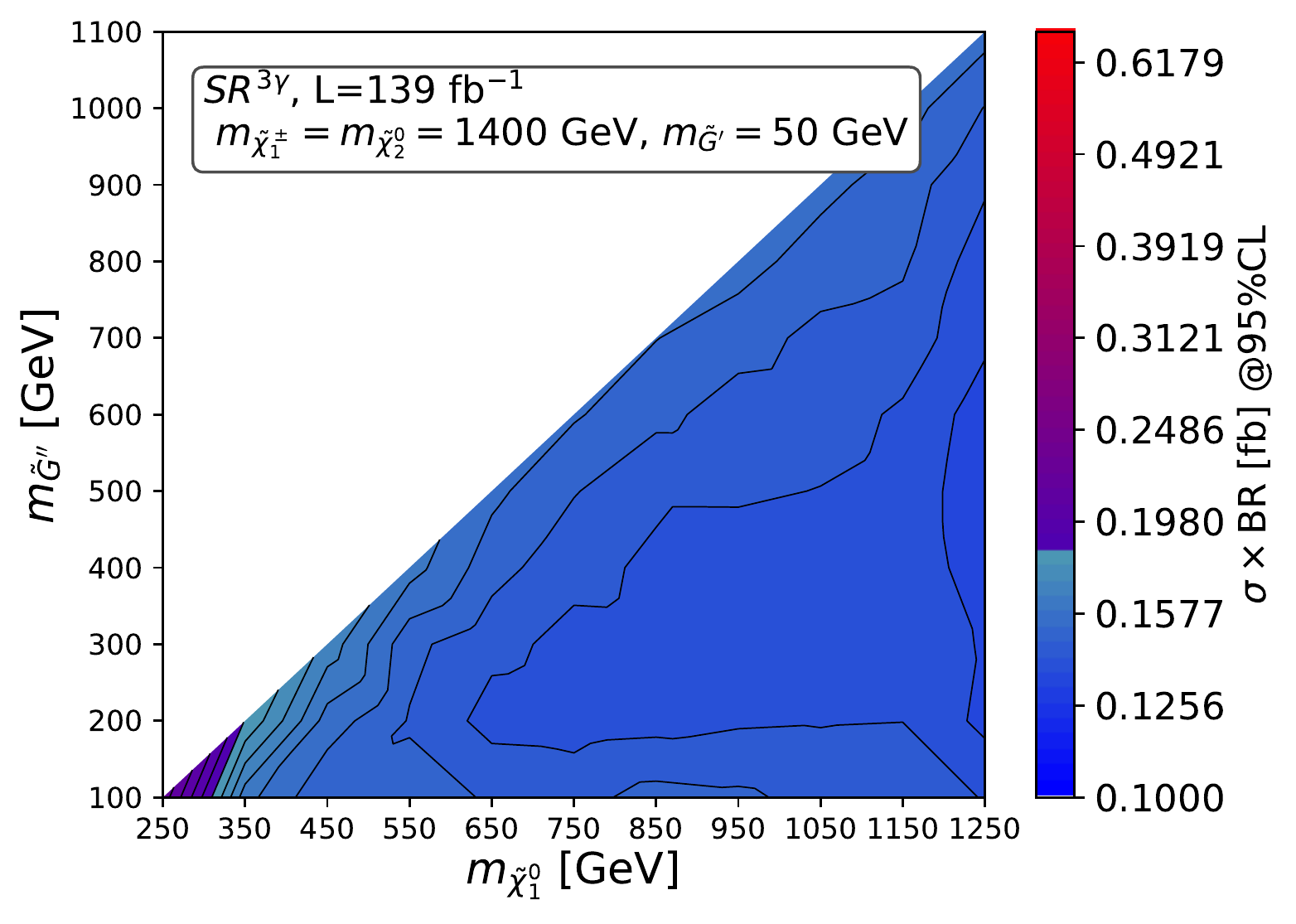}
\includegraphics[width=0.45\textwidth]{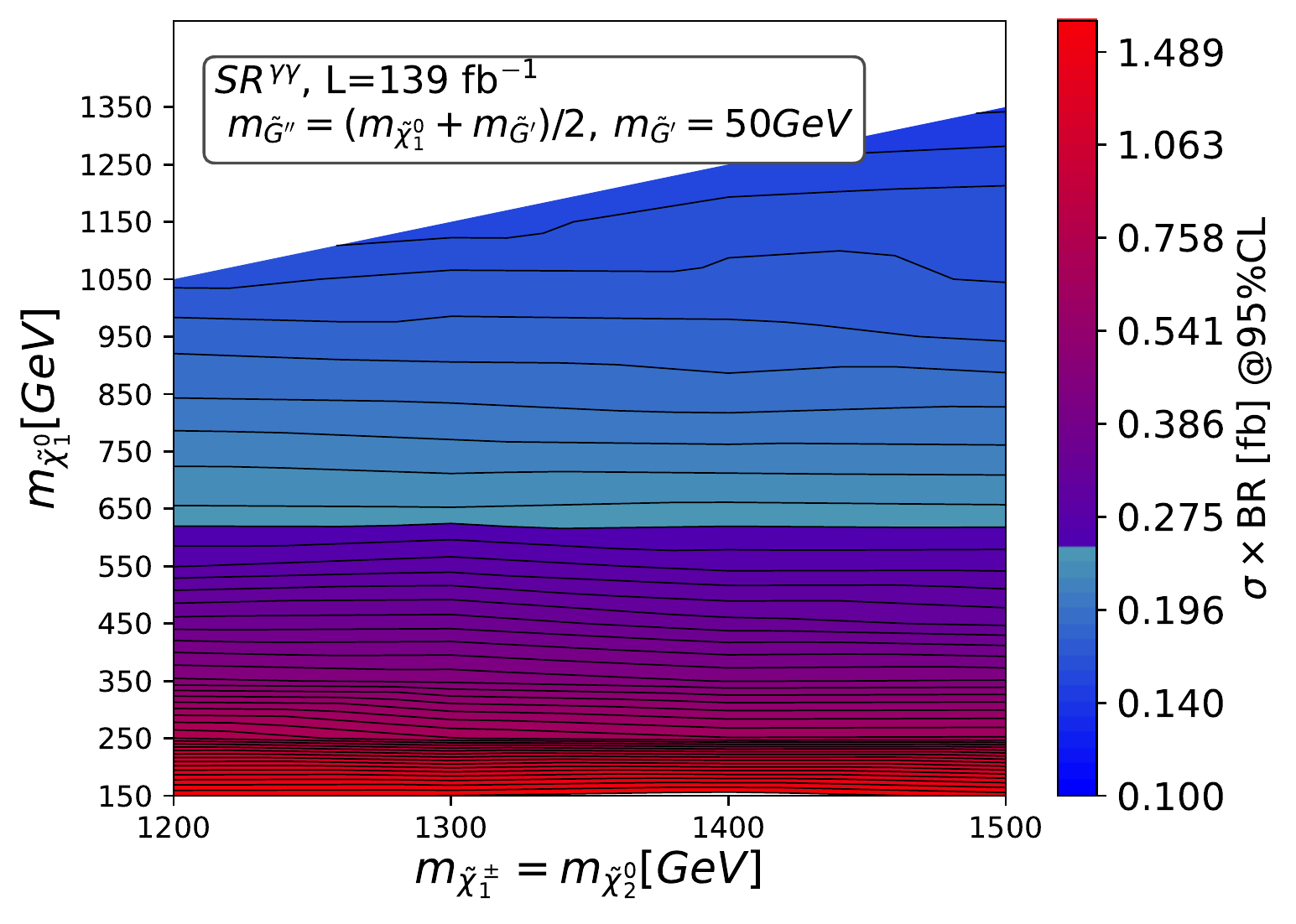}
\includegraphics[width=0.45\textwidth]{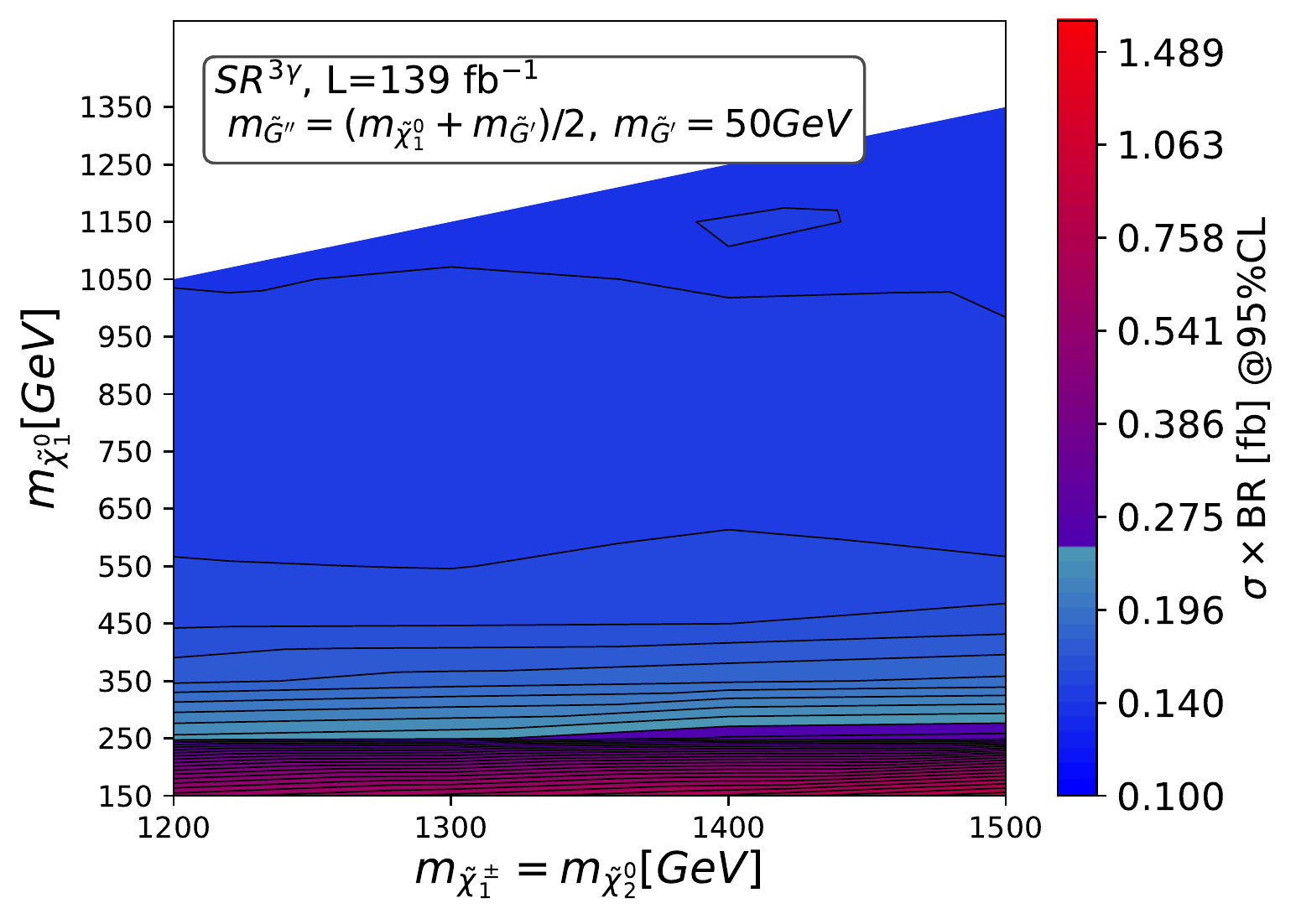}
\includegraphics[width=0.45\textwidth]{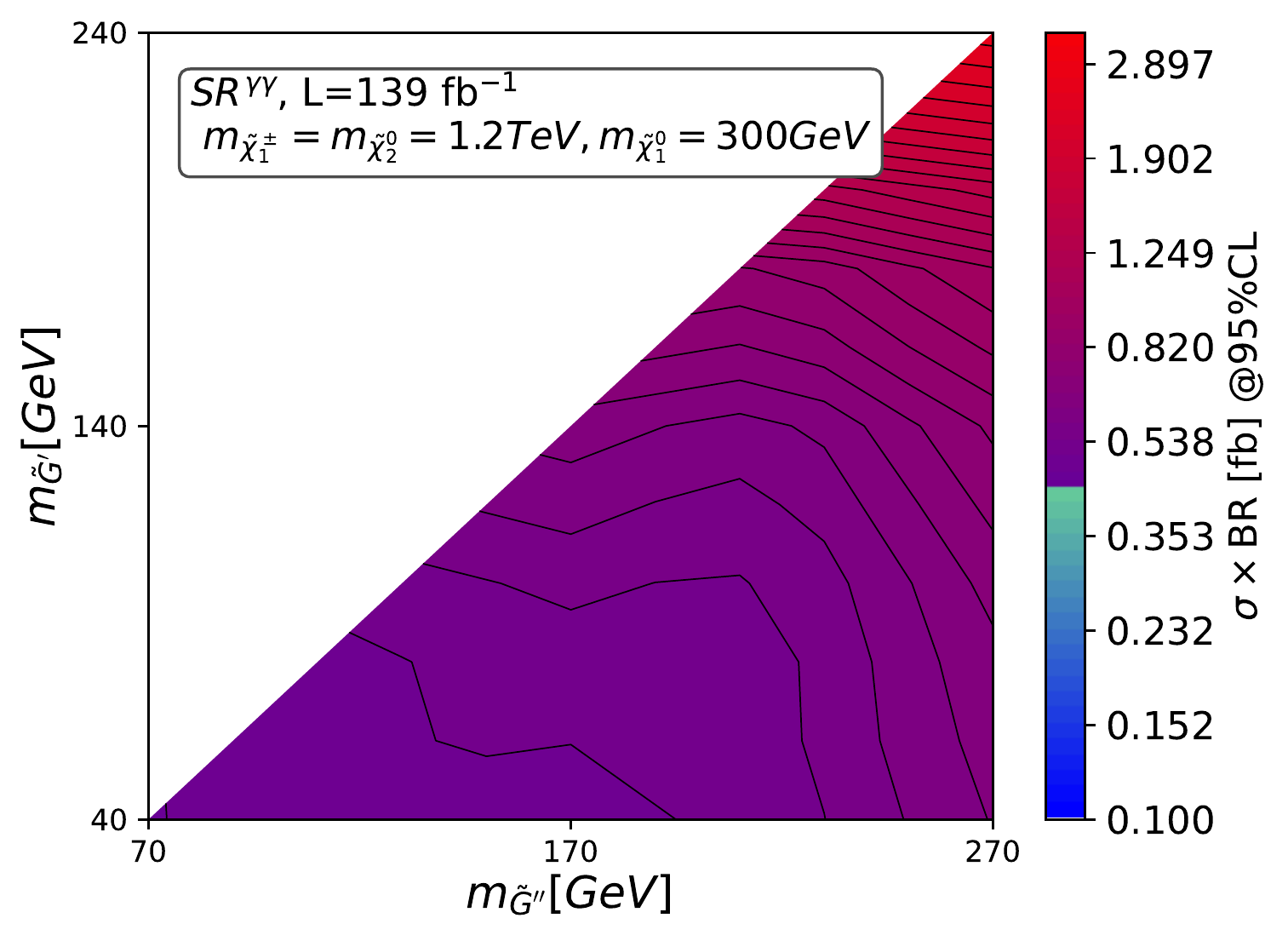}
\includegraphics[width=0.45\textwidth]{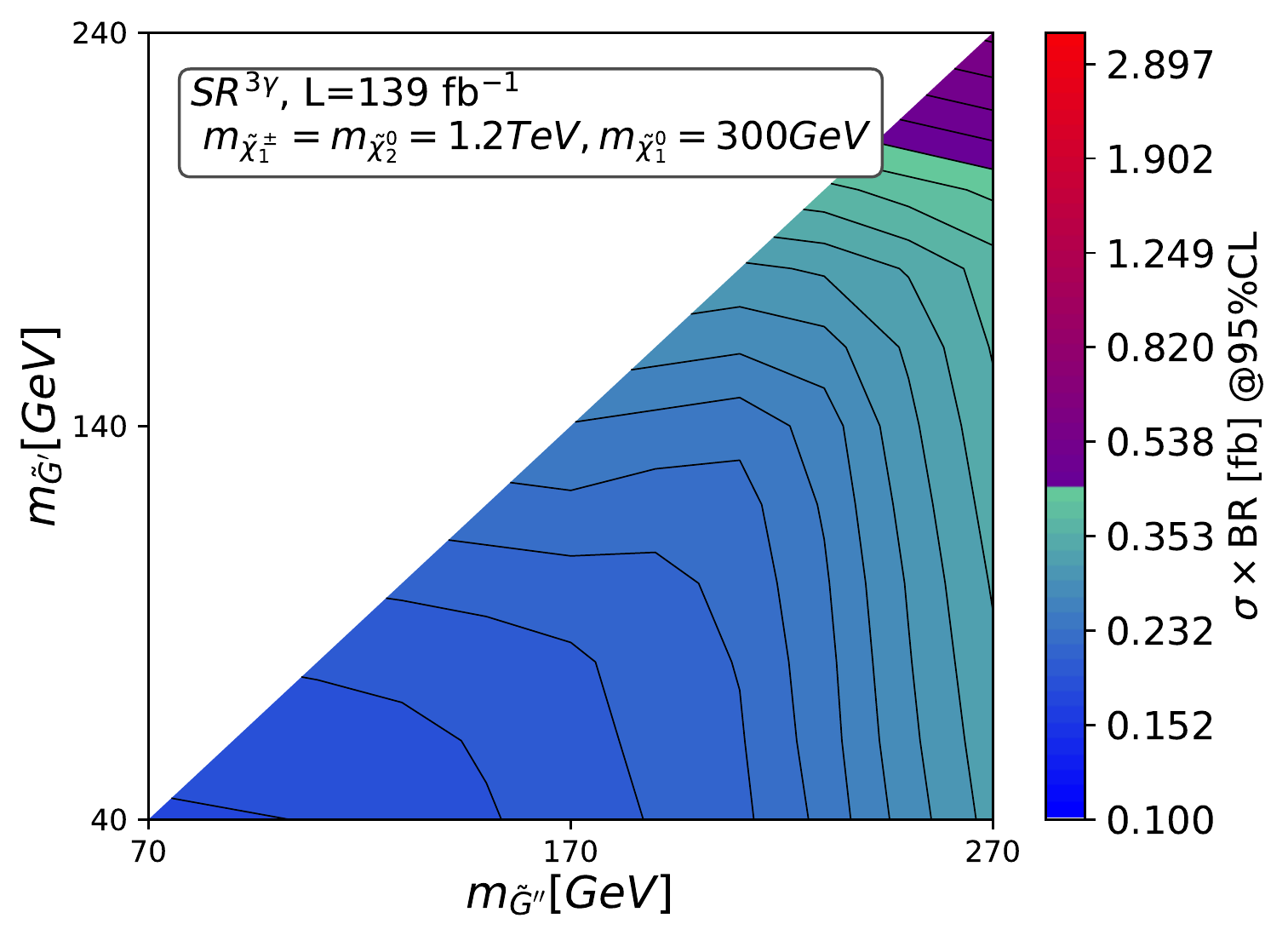}
\caption{Upper limits at 95\% C.L. on the production cross section times branching ratio (in fb) in the complementary grids from \SRgg (left) and \SRggg (right). The limits are derived for an integrated luminosity of $\mathcal{L}=139\ifb$.
}
\label{fig:bounds_SRWH}
\end{figure}

For each of the benchmark points described in Sec.~\ref{sec:benchmarks}, an exclusion test is carried out
where the upper limit on the signal cross section from either \SRgg or \SRggg is compared to the production cross section of the benchmark point. The results are shown in \fig{fig:bounds_models}. Black circles depict benchmark points excluded by both \SRgg and \SRggg, the few green crosses are excluded only by \SRgg, red squares are excluded only by \SRggg and blue triangles are not excluded by either of the two signal regions. It can be noted that a large set of benchmark points are only excluded by \SRggg and not by \SRgg. As these are realistic models, several processes contribute to any given benchmark point. Hence to obtain the upper limits we compute the cross sections for the various processes and multiply these with the corresponding selection efficiencies $\epsilon_{\gamma\gamma}$ or $\epsilon_{3\gamma}$. In the case of the main process, the efficiency from the closest point in the chargino-neutralino grid is used for a given benchmark point. For each of the secondary processes the ratio of efficiencies with respect to the main process is calculated for the following points in the chargino-neutralino grid: ($m_{\ninotwo}=m_{\chinoonepm}[\text{GeV}]$,$\, m_{\ninoone}[\text{GeV}]$) = ($800,500$), ($900,700$), ($1300,700$), ($1400,900$), ($1400,1100$). The averages of those ratios across the five grid points are given in Tab.~\ref{tab:efficiencies} and are used to scale the efficiency of the main process to obtain approximate efficiencies for the secondary processes in a given benchmark point. For $\epsilon_{\gamma\gamma}$, the relative variation of the ratios across the five grid points is below 11\% for all processes. In the case of $\epsilon_{3\gamma}$ the relative variation is below 3\% for all processes with three or more photons. The variation for processes with two photons is larger, but these have a negligible contribution to the overall efficiency.

\begin{table}[htbp]
    \centering
    \begin{tabular}{c|c|c}
Process &    	Ratio $\epsilon_{\gamma\gamma}$ & Ratio $\epsilon_{3\gamma}$ \\
\hline
$pp\to W^\pm\,h\, 4\gamma\, \pgldtwo\,\pgldtwo$	& 1   & 1 \\
$pp\to W^+\,W^-\, 4\gamma\, \pgldtwo\,\pgldtwo$	& $1.077 \pm 0.052$  & $1.051 \pm 0.026$\\
$pp\to W^\pm\,h\, 3\gamma\, \pgldtwo\,\pgldtwo$	& $1.174 \pm 0.100$ & $0.634 \pm 0.016$\\
$pp\to W^+\,W^-\, 3\gamma\, \pgldtwo\,\pgldtwo$	& $1.207 \pm 0.072$	& $0.678 \pm 0.017$\\
$pp\to W^\pm\,h\, 2\gamma\, \pgldtwo\,\pgldtwo$	& $0.972 \pm 0.108$	& $0.038 \pm 0.017$\\
$pp\to W^+\,W^-\, 2\gamma\, \pgldtwo\,\pgldtwo$	& $1.003 \pm 0.088$	& $0.053 \pm 0.013$\\
    \end{tabular}
    \caption{Average ratio of efficiencies for each process with respect to the main process \ref{eq:dominantproc} (first line) for \SRgg and \SRggg. The uncertainties are the standard deviations of the ratios across the five grid points from which the averages have been obtained.}
    \label{tab:efficiencies}
\end{table}

\begin{figure}[htbp]
\centering
\includegraphics[width=0.49\textwidth]{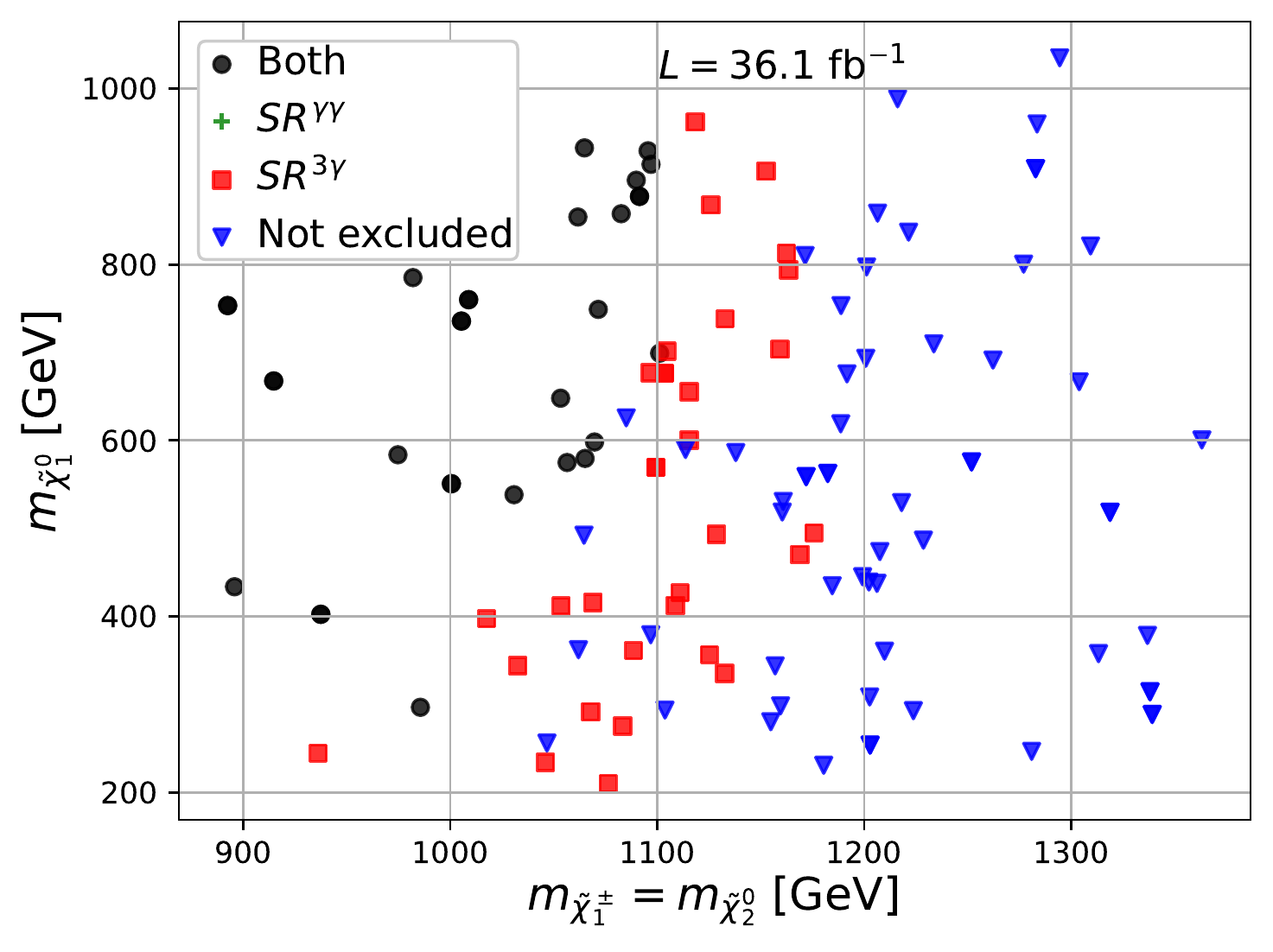}
\includegraphics[width=0.49\textwidth]{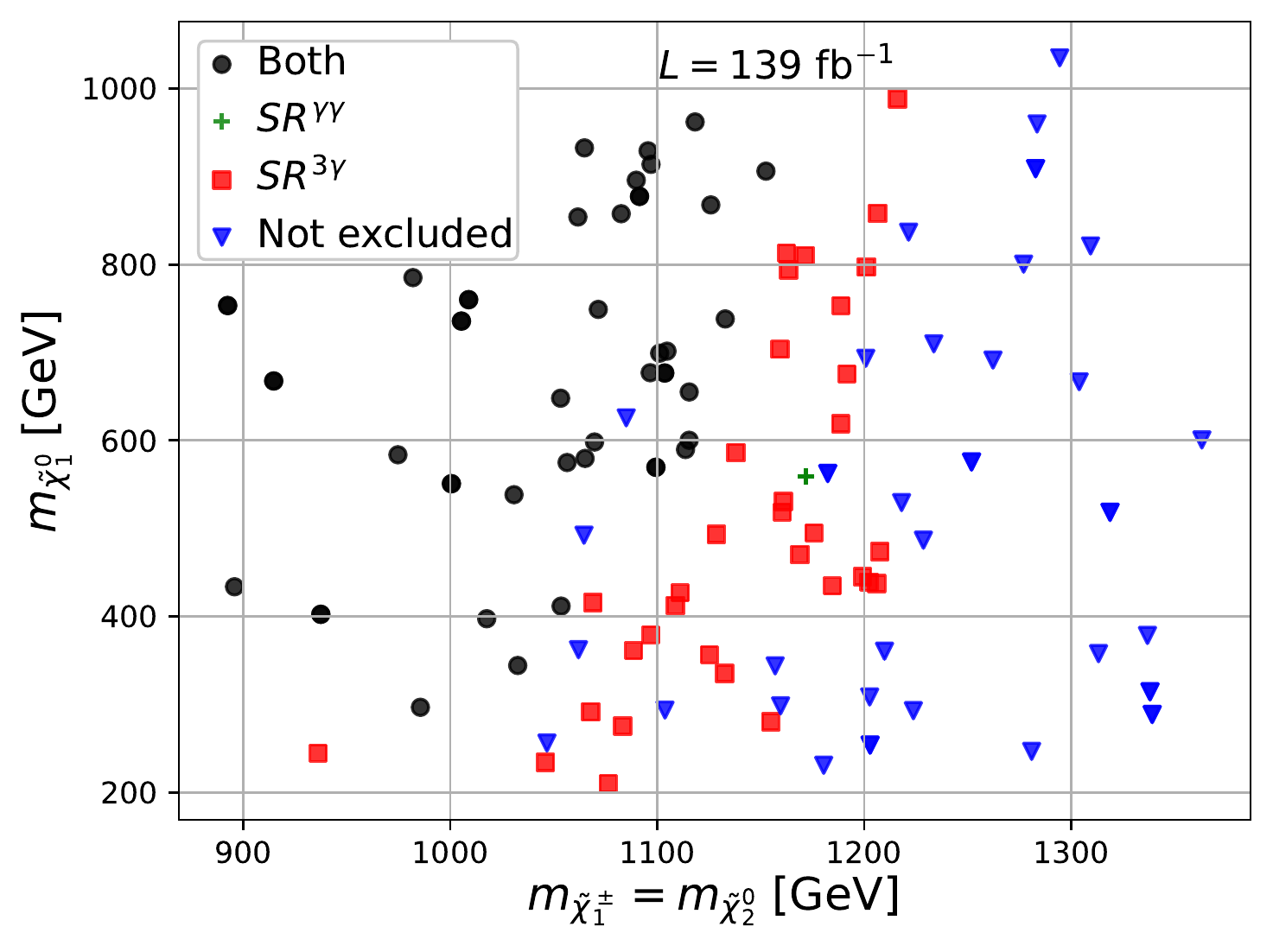}
\includegraphics[width=0.49\textwidth]{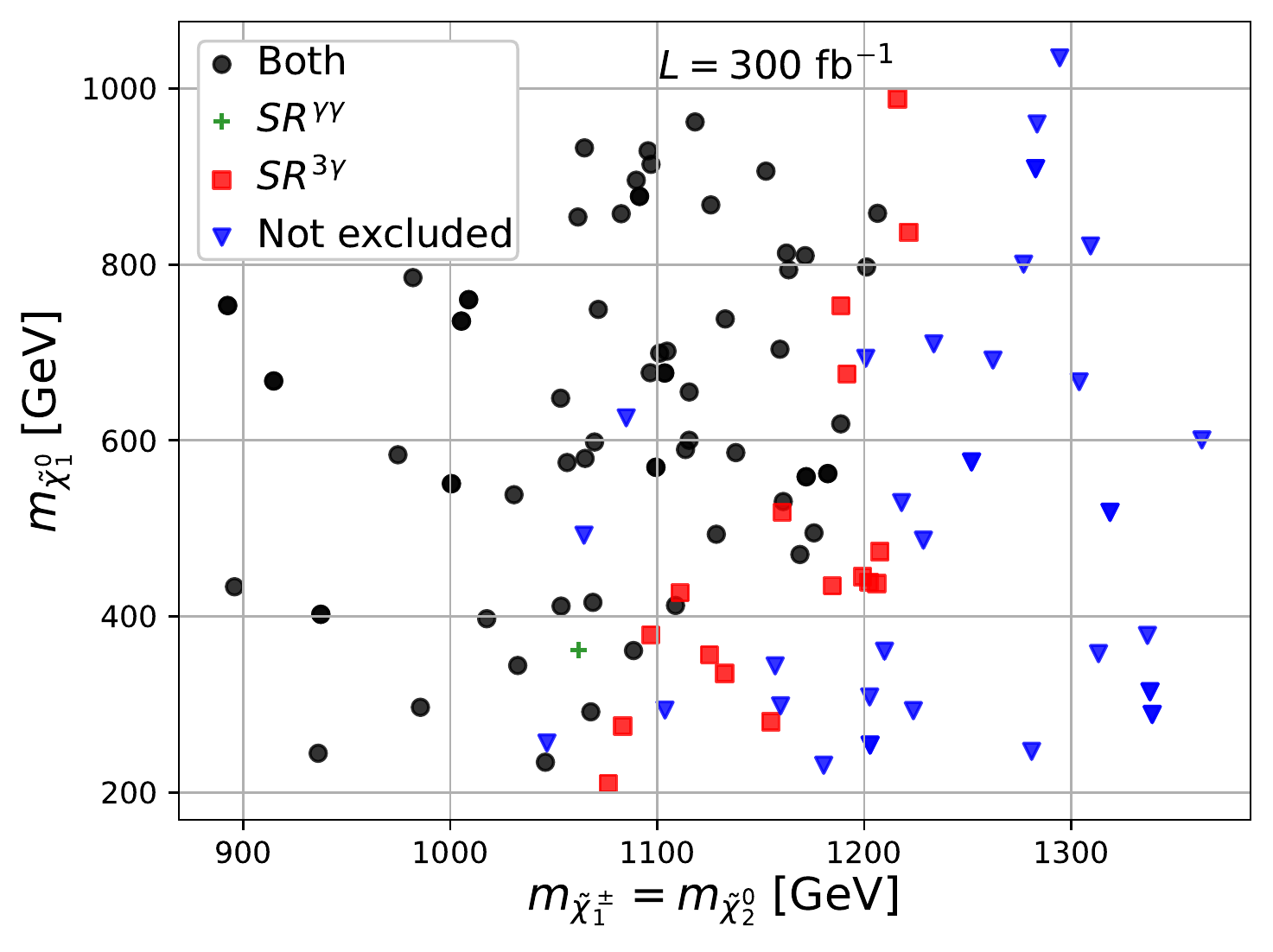}
\includegraphics[width=0.49\textwidth]{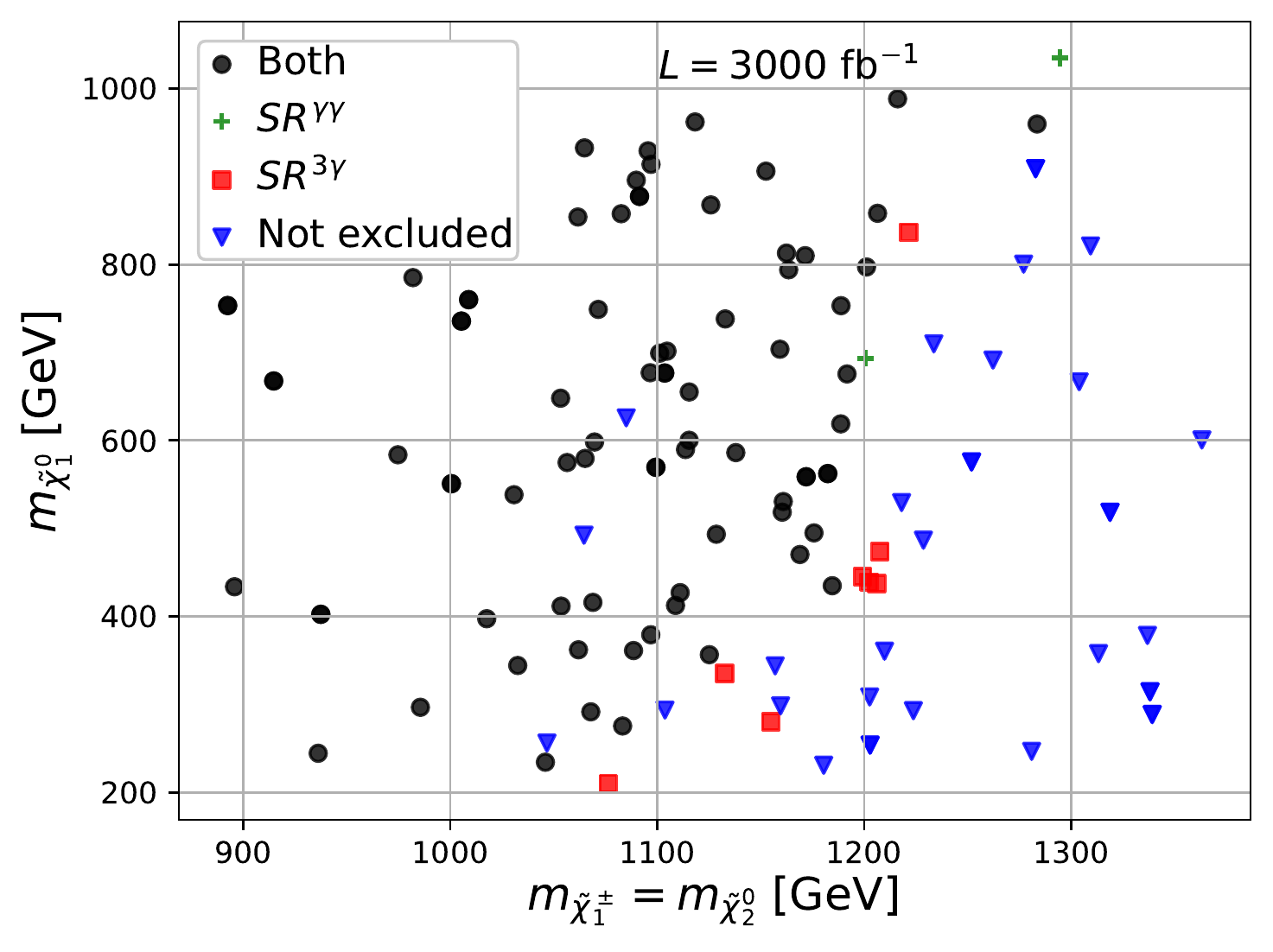}
\caption{Exclusion results for the benchmark points from Sec.~\ref{sec:benchmarks} for the four luminosity scenarios. Black circles depict benchmark points excluded by both \SRgg and \SRggg, green crosses are excluded only by \SRgg, red squares are excluded only by \SRggg and blue triangles are excluded neither by \SRgg nor \SRggg.}
\label{fig:bounds_models}
\end{figure}

\clearpage

\section{Conclusions}
\label{conclusion}

This paper presents studies of the NMSSM with multiple sectors of gauge-mediated supersymmetry breaking and strategies to search for such models at the Large Hadron Collider. The models considered are restricted to those with three additional sectors and the collider signatures studied involve at least three photons. The production of \chinoonepm and \ninotwo, both decaying to the collider-stable \pgldtwo while emitting two photons and a $W$ or Higgs boson is used as a benchmark process. Limits are placed on the production cross section times branching ratio in a grid spanning the $\chinoonepm=\ninotwo$ and \ninoone masses with the mass of the \pgldthree fixed at 80\,GeV and the mass of the \pgldtwo fixed at 50\,GeV. The most stringent limits are achieved when requiring at least three isolated photons, 50\,GeV of missing transverse energy, \met, and 400\,GeV of scalar transverse energy, \HT.

The three-photon analysis shows better cross-section limits than the two-photon analysis across the entire scanned mass plane. In particular, the two-photon analysis shows significantly weaker limits at low chargino and neutralino masses. For an integrated luminosity of 139~\ifb, the upper limit on the cross section times branching ratio from the three-photon analysis is $0.32\,\fb$ for the grid point ($m_{\ninotwo}=m_{\chinoonepm}[\text{GeV}]$,$\, m_{\ninoone}[\text{GeV}]$) = ($500,200$)
and reaches $0.14\,\fb$ for the grid point
($m_{\ninotwo}=m_{\chinoonepm}[\text{GeV}]$,$\, m_{\ninoone}[\text{GeV}]$) = ($1500,1350$).
For the two-photon analysis the corresponding limits are $14.8\,\fb$ and $0.19\,\fb$. The limits are also found to vary with the masses of the \pgldtwo and \pgldthree, but in all regions explored the three-photon analysis outperforms the two-photon analysis.

The cross section limits are also applied to the set of benchmark points found by scanning the parameter space of the NMSSM with multiple sectors. The three-photon analysis is found to exclude a larger set of points than the two-photon analysis carried out in~\cite{Aaboud:2018doq}. For all \ninoone masses, the three-photon analysis outperforms the two-photon analysis and the difference in performance tends to grow towards lower \ninoone masses.

\section*{Acknowledgments}
This work is supported by the Knut and Alice Wallenberg foundation under the grant KAW 2017.0100 (SHIFT project).
We would like to thank R. Argurio and A. Mariotti for comments on the draft of the manuscript.
Computational resources have been provided by the supercomputing facilities of the Université catholique de Louvain (CISM/UCL) and the Consortium des Équipements de Calcul Intensif en Fédération Wallonie Bruxelles (CÉCI) funded by the Fond de la Recherche Scientifique de Belgique (F.R.S.-FNRS) under convention 2.5020.11 and by the Walloon Region.

\appendix

\section{Comparison with existing two-photon ATLAS search}
\label{sec:comparison}

In this Appendix the predictions of the ATLAS search with a diphoton and \met signature in~\cite{Aaboud:2018doq}, referred to as the diphoton search, are compared to those in this paper in order to validate the analysis method used.

In the signal region \SRgg (Eq.~\ref{eq:SRgg})~\cite{Aaboud:2018doq} predicts $2.05^{+0.65}_{-0.63}$ SM background events
for $\mathcal{L}=36.1\ifb$, resulting in an expected model-independent upper limit on the visible cross section for non-SM processes of 
$\sigma_{\text{exp}}^{95} = 0.122\fb$
(see Tab.~\ref{tab:bounds}).
Using the simplified statistical model described in Sec.~\ref{sec:results} with $z=2$ we obtain an expected upper limit $\sigma_{\text{exp}}^{95} =0.113\fb$.

In addition to comparing the cross-section limits, our selection efficiencies for \SRgg ($\epsilon_{\gamma\gamma}$) are compared to those in \cite{Aaboud:2018doq} ($\epsilon_{\footnotesize{\rm ATLAS}}$\footnote{\url{https://www.hepdata.net/record/ins1654357?version=1&table=Acceptance\%2FEfficiency\%206}}). We do this by generating events for the following process
\begin{equation}
pp\to (\chinoonepm \to W^\pm\,(\ninoone \to \tilde{G}\, \gamma)) 
    + (\ninotwo \to h\,(\ninoone \to \tilde{G}\, \gamma))
\label{eq:diphoton_process}
\end{equation}
in a model with one SUSY-breaking sector and a massless gravitino. The efficiencies are listed in Tab.~\ref{tab:efficiencies_compare} for four different mass points.
Since in~\cite{Aaboud:2018doq} the bino-like $\tilde\chi^0_1$ decays to the \pgldone by emitting either a photon or a $Z$ boson, where BR$(\ninoone \to \pgldone \gamma)\sim \cos^2\theta_W$, the same is assumed for our generated signal samples. 
\begin{table}[htbp]
    \centering
    \begin{tabular}{c|c|c}
$(m_{\chinoonepm}[\text{GeV}],m_{\ninoone}[\text{GeV}])$ & $\epsilon_{\footnotesize{\rm ATLAS}}$ & $\epsilon_{\gamma\gamma}$ \\
\hline
$(800,200)$ & 0.066 & 0.051 \\
$(900,500)$ & 0.165 &  0.130\\
$(1000,400)$ & 0.151 & 0.131\\
$(1000,800)$ & 0.207 & 0.154\\
    \end{tabular}
    \caption{Selection efficiencies from \cite{Aaboud:2018doq}, $\epsilon_{\footnotesize{\rm ATLAS}}$, and from this paper, $\epsilon_{\gamma\gamma}$, for different mass points.}
    \label{tab:efficiencies_compare}
\end{table}
The discrepancy between the selection efficiencies is expected since the results presented in this paper are based on a simplified detector simulation, neglect the effects of pileup and use jets reconstructed with a distance parameter $R=0.6$ rather than the $R=0.4$ used by the ATLAS analysis. Moreover, signal samples in which the $\ninoone$ decays by emitting a $Z$ boson rather than a photon have been omitted.

To validate our background estimate, our prediction $N=0.77\pm 0.27$ for $W(\ell\nu)\gamma\gamma$ is compared to the corresponding $N = 1.08^{+0.65}_{-0.63}$ from \cite{Aaboud:2018doq}.
The quoted uncertainty includes both statistical and systematic effects, so the central value of the prediction differ by $\sim 40\%$. This is expected since NNLO corrections are large at large $H_T$~\cite{Gehrmann-DeRidder:2018kng}.


\bibliography{GMbib}
\bibliographystyle{JHEP}
\end{document}